\documentclass{aa}  
\usepackage{graphicx}
\usepackage{txfonts}
\usepackage{hyperref}
\usepackage{pack/multirow}
\usepackage{pack/soul}
\usepackage{pack/float}
\usepackage{pack/gensymb}
\usepackage{pack/color}
\usepackage{pack/threeparttable}
\usepackage{verbatim}

\definecolor{cadmiumgreen}{rgb}{0.0, 0.42, 0.24}
\definecolor{caribbeangreen}{rgb}{0.0, 0.8, 0.6}
\definecolor{darkcerulean}{rgb}{0.03, 0.27, 0.49}
\definecolor{darkcyan}{rgb}{0.0, 0.55, 0.55}
\definecolor{darkpastelgreen}{rgb}{0.01, 0.75, 0.24}
\definecolor{amber(sae/ece)}{rgb}{1.0, 0.49, 0.0}

\newcommand{\orcid}[1]{\href{https://orcid.org/#1}{\includegraphics[width=10pt]{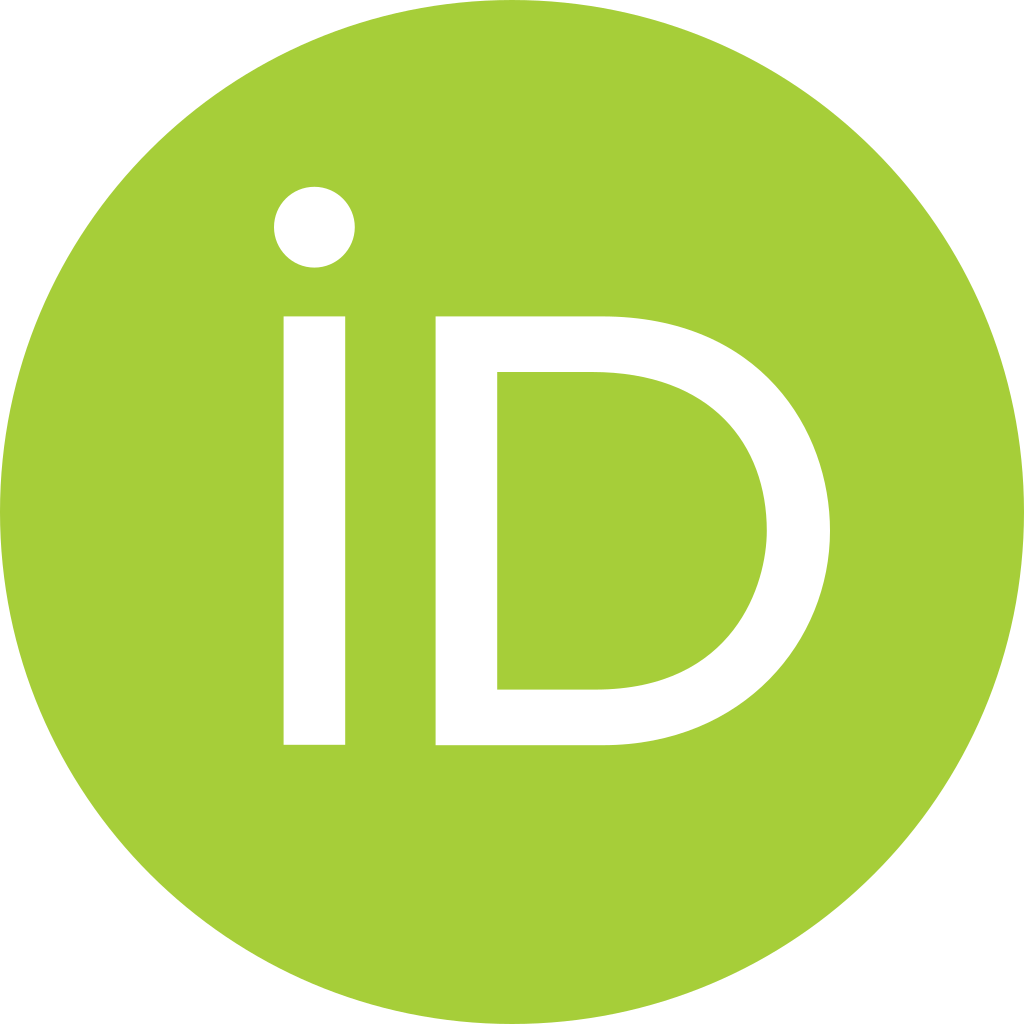}}}

\begin{document} 

   \title{The VANDELS ESO public spectroscopic survey: \\ The spectroscopic measurements catalogue\thanks{The measurement catalogues are accessible through the survey database (\url{http://vandels.inaf.it}) where all information can be queried interactively, and at the CDS via anonymous ftp to \url{cdsarc.cds.unistra.fr} (130.79.128.5) or via \url{https://cdsarc.cds.unistra.fr/cgi-bin/qcat?J/A+A/}.}}
   \author{M. Talia\orcid{0000-0003-4352-2063}\inst{1,2},
                C. Schreiber,
                B. Garilli\orcid{0000-0001-7455-8750}\inst{3},
                L. Pentericci\orcid{0000-0001-8940-6768}\inst{4},
                L. Pozzetti\inst{2},
                G. Zamorani\inst{2},
                F. Cullen\orcid{0000-0002-3736-476X}\inst{5},
                M. Moresco\orcid{0000-0002-7616-7136}\inst{1,2},
                A. Calabrò\orcid{0000-0003-2536-1614}\inst{4},
                M. Castellano\orcid{0000-0001-9875-8263}\inst{4},
                J. P. U. Fynbo\inst{6,7},
                L. Guaita\orcid{0000-0002-4902-0075}\inst{8},
                F. Marchi,
                S. Mascia\orcid{0000-0002-9572-7813}\inst{4,9},
                R. McLure\inst{5},
                M. Mignoli\inst{2},
                E. Pompei\inst{10},
                E. Vanzella\inst{2},
                A. Bongiorno\inst{11},
                G. Vietri\orcid{0000-0001-9155-8875}\inst{4,3},
                R. O. Amor\'{i}n\orcid{0000-0001-5758-1000}\inst{11,12},
                M. Bolzonella\inst{2},
                A. C. Carnall\orcid{0000-0002-1482-5818}\inst{5},
                A. Cimatti\orcid{0000-0002-4409-5633}\inst{1,2},
                G. Cresci\orcid{0000-0002-5281-1417}\inst{13},
                S. Cristiani\inst{14,15},
                O. Cucciati\orcid{0000-0002-9336-7551}\inst{2},
                J. S. Dunlop\inst{5},
                F. Fontanot\orcid{0000-0003-4744-0188}\inst{14,15},
                P. Franzetti\orcid{0000-0002-6986-0127}\inst{3},
                A. Gargiulo\orcid{0000-0002-3351-1216}\inst{3},
                M. L. Hamadouche\orcid{0000-0001-6763-5551}\inst{5},
                N. P. Hathi\orcid{0000-0001-6145-5090}\inst{16},
                P. Hibon\inst{10},
                A. Iovino\inst{17},
                A. M. Koekemoer\orcid{0000-0002-6610-2048}\inst{16},
                F. Mannucci\orcid{0000-0002-4803-2381}\inst{13},
                D. J. McLeod\inst{5},
                \and
                A. Saldana-Lopez\orcid{0000-0001-8419-3062}\inst{18}
        }
   \institute{
        \textit{University of Bologna - Department of Physics and Astronomy “Augusto Righi” (DIFA), Via Gobetti 93/2, I-40129 Bologna, Italy}\\
   \email{margherita.talia2@unibo.it}
   \and 
   \textit{INAF-Osservatorio di Astrofisica e Scienza dello Spazio, Via Gobetti 93/3, I-40129, Bologna, Italy}
   \and 
   \textit{INAF-IASF Milano, Via Alfonso Corti 12, I-20133 Milano, Italy}
   \and 
   \textit{INAF-Osservatorio Astronomico di Roma, via Frascati 33, 00078, Monteporzio Catone, Italy}
   \and 
   \textit{Institute for Astronomy, University of Edinburgh, Royal Observatory, Edinburgh EH9 3HJ, UK}
   \and 
   \textit{Cosmic Dawn Center (DAWN), Copenhagen, Denmark}
   \and 
   \textit{Niels Bohr Institute, University of Copenhagen, Jagtvej 128, 2200 Copenhagen N, Denmark}
   \and 
   \textit{Departamento de Ciencias Fisicas, Facultad de Ciencias Exactas, Universidad Andres Bello, Fernandez Concha 700, Las Condes, Santiago, Chile}
   \and 
   \textit{Dipartimento di Fisica, Università di Roma Tor Vergata, Via della Ricerca Scientifica, 1, 00133, Roma, Italy}
   \and 
   \textit{European Southern Observatory, Alonso de Córdova 3107, Vitacura, Santiago de Chile, Chile}
   \and 
   \textit{Instituto de Investigaci\'{o}n Multidisciplinar en Ciencia y Tecnolog\'{i}a, Universidad de La Serena, Raul Bitr\'{a}n 1305, La Serena 2204000, Chile}
   \and 
   \textit{Departamento de Astronom\'{i}a, Universidad de La Serena, Av. Juan Cisternas 1200 Norte, La Serena 1720236, Chile}
   \and 
   \textit{INAF-Osservatorio Astrofisco di Arcetri, largo E. Fermi 5, 50127 Firenze, Italy}
   \and 
   \textit{INAF-Astronomical Observatory of Trieste, via G.B. Tiepolo 11, I-34143 Trieste, Italy}
   \and 
   \textit{IFPU-Institute for Fundamental Physics of the Universe, via Beirut 2, I-34151 Trieste, Italy}
   \and 
   \textit{Space Telescope Science Institute, 3700 San Martin Dr.,
   Baltimore, MD 21218, USA}
   \and 
   \textit{INAF - Osservatorio Astronomico di Brera, via Brera 28, I-20121
Milano, Italy}
   \and 
   \textit{Department of Astronomy, University of Geneva, 51 Chemin Pegasi, 1290 Versoix, Switzerland}
   }

   \abstract{VANDELS is a deep spectroscopic survey, performed with the VIMOS  instrument at VLT, aimed at studying in detail the physical properties of high-redshift galaxies. VANDELS targeted $\sim$2100 sources at $1<z<6.5$ in the CANDELS Chandra Deep-Field South (CDFS) and Ultra-Deep Survey (UDS) fields.  
   In this paper we present the public release of the spectroscopic measurement catalogues from this survey, featuring emission and absorption line centroids, fluxes, and rest-frame equivalent widths obtained through a Gaussian fit, as well as a number of atomic and molecular indices (e.g. Lick) and continuum breaks (e.g. D4000), and including a correction to be applied to the error spectra.  
   We describe the measurement methods and the validation of the codes that were used. 
   }

   \authorrunning{M. Talia et al.} 
   \titlerunning{VANDELS spec measu}

   \date{Received ; accepted }
   \keywords{ surveys - catalogs }
   \maketitle
%
\section{Introduction}\label{sec:introduction}
A major theme in extragalactic astronomy is understanding when and how galaxies formed and evolved. 
Spectroscopic surveys play a fundamental role in this respect, not only because they provide robust redshifts, but especially because the analysis of emission and absorption lines and spectral breaks grants access to intrinsic physical properties of galaxies such as the chemical composition of their gas and stellar populations, the ionising radiation field, and the gas and star kinematics.

Over the past two decades, several multi-slit and multi-fibre surveys have been carried out, targeting increasingly distant galaxies: from the 
Sloan Digital Sky Survey (SDSS) in the local Universe \citep{abazajian2003,abdurrouf2022}, passing through the VIMOS VLT\footnote{VIMOS = Visible Multi Object Spectrograph; VLT = Very Large Telescope} Deep Survey \citep[VVDS; ][]{lefevre2013, garilli2008}, zCOSMOS \citep{lilly2007}, VIMOS Public Extragalactic Redshift Survey 
\citep[VIPERS; ][]{guzzo2014,scodeggio2018}, and the Large Early Galaxy Census \citep[LEGA-C; ][]{vanderwel2016} at $<z>\sim 0.7$, the Galaxy Mass Assembly ultradeep Spectroscopic Survey \citep[GMASS; ][]{cimatti2008,kurk2013} at Cosmic Noon, and up to $z\sim 4-6$ with KBSS-MOSFIRE\footnote{KBSS = Keck Baryonic Structure Survey} \citep{steidel2014} and the VIMOS Ultra Deep Survey \citep[VUDS; ][]{lefevre2015}, along with a number of smaller samples targeting the re-ionisation epoch \citep[e.g.][]{pentericci2018a}.
All these surveys have improved our understanding of galaxy evolution, mainly by drawing a detailed 3D map of the Universe with thousands of redshifts.

VANDELS is an ESO public VIMOS survey of the Chandra Deep-Field South (CDFS) and Ultra-Deep Survey (UDS) fields that was designed to complement and extend the work of the CANDELS \citep{grogin2011, koekemoer2011} imaging campaigns.
The strategy of VANDELS was not to limit itself to finding a redshift, but to focus on ultra-long exposures of a relatively small number of galaxies that provide high signal-to-noise ratio (S/N) spectra to study in detail the physical characteristics of the high-redshift galaxies \citep{mclure2018}. 
Since the first data release of VANDELS \citep{pentericci2018}, a number of papers have been published studying several properties, ranging from 
dust attenuation, interstellar medium properties, and stellar metallicities of star-forming \citep{cullen2018,cullen2019,calabro2021,fontanot2021,calabro2022a,calabro2022b} 
and quiescent galaxies \citep{carnall2019,carnall2020,carnall2022,hamadouche2022, tomasetti2023, hamadouche2023}, 
to intergalactic medium properties \citep{thomas2020,thomas2021}, 
the ionising photon production efficiency \citep{castellano2023},
the LyC escape fraction \citep{begley2022,saldanalopez2022}, 
Ly$\alpha$, HeII$\lambda$1640, CIV$\lambda$1550, and CIII]$\lambda$1908 emitters \citep{marchi2019,hoag2019,cullen2020,saxena2020a,saxena2020b,guaita2020,saxena2022,llerena2022,mascia2023}, 
AGN \citep{magliocchetti2020}, 
and high-mass X-ray binaries \citep{saxena2021}.

This paper represents the official release of the the VANDELS spectroscopic measurements (i.e. lines, indices, and breaks), which are herewith made available to the whole astrophysical community.
The catalogues include all spectra from the VANDELS final data release (DR4) presented in \citet{garilli2021} with a robust spectroscopic redshift.

The paper is organised as follows: Section 2 briefly describes the VANDELS survey; Section~3 summarises the methods of measurements; Section 4 discusses the measurement code validation tests, including a description of the creation of ad hoc mock 1D spectra; Section 5 discusses an issue with the error spectra and its resolution; Section 6 describes the released catalogues and the comparison with independent measurements from previously published works inside the VANDELS collaboration; and Section 7 provides a brief summary.
In this paper, we provide magnitudes in the AB photometric system \citep{oke1983}.

        \begin{figure}
        \centering
        \includegraphics[scale=0.5]{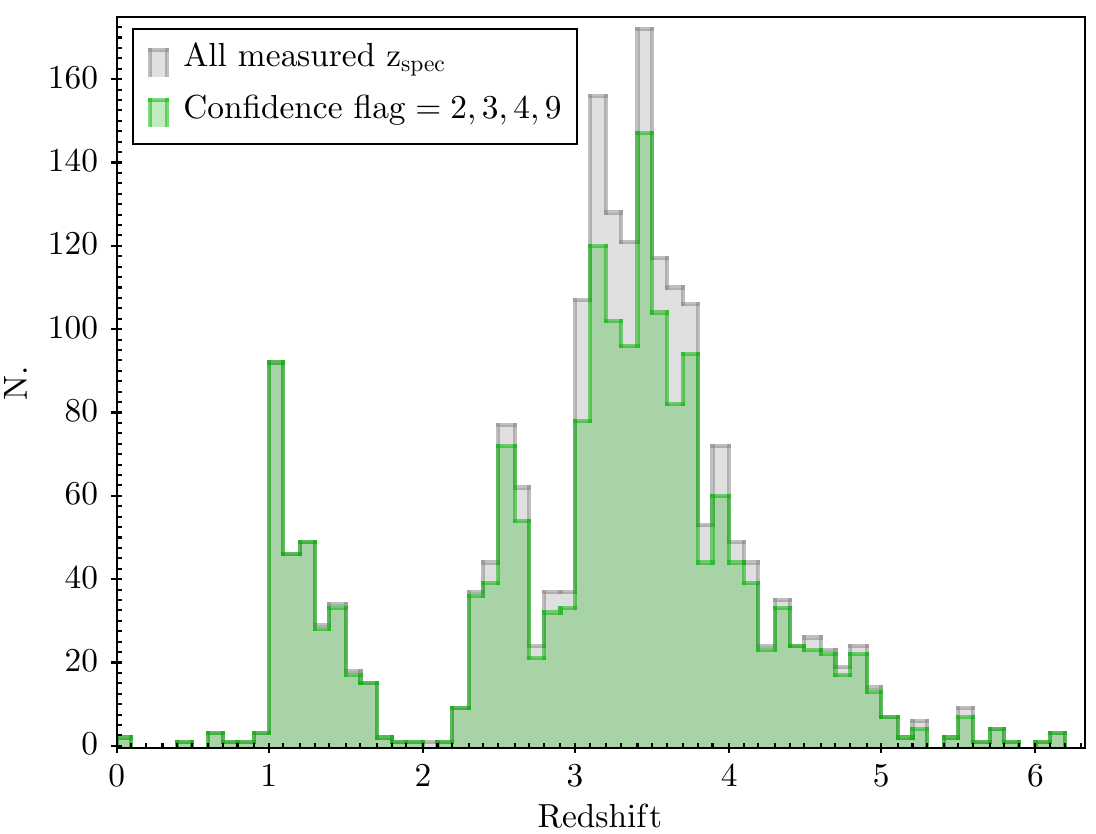}
        \caption{Redshift distribution of the final VANDELS sample: the grey histogram includes all measurements, and the green histogram includes only reliable redshifts (confidence flag > 1; see the text for more details).}
        \label{zhist}
        \end{figure}

\section{The VANDELS survey}\label{sec:vandels}
VANDELS is a spectroscopic survey performed with the ESO-VLT VIMOS spectrograph in two CANDELS fields, over a total area of $\sim$ 0.2 square degree. 
For all the details on the survey design, target selection, observations, data reduction, and spectroscopic redshift measurements, we refer the reader to \citet{mclure2018,pentericci2018,garilli2021}.

The VANDELS spectra cover a wavelength range of 4800~\AA $< \lambda_{obs} <$ 9800~\AA, with a dispersion of 2.5~\AA/pixel and a spectral resolution of $R \sim 650$, corresponding to a FWHM$_{res}\sim$ 460 km/s (or FWHM$_{res}\sim$ 11.2~\AA~ at 7300~\AA).
The main targets of the survey were massive passive galaxies at $1 < z < 2.5$, bright star-forming galaxies (SFGs) at $2.4 < z < 5.5$, and fainter SFGs at $3 < z < 7$ Lyman-break galaxies, plus a small sample of AGN, pre-selected using various multi-wavelength criteria. 
The VANDELS spectroscopic targets were pre-selected using high-quality photometric redshifts and were observed for a minimum of 20 hrs and up to 80 hrs, depending on their brightness, in order to ensure an approximately homogeneous S/N on the continuum within each class of galaxy. 
The data reduction was carried out using the recipes provided by the VIMOS Interactive Pipeline and Graphical Interface (\texttt{VIPGI}) package \citep{scodeggio2005} and the \texttt{EASYLIFE} environment \citep{garilli2012}.
The measured S/N per resolution element is higher than ten 
for all passive 
and star-forming galaxies, and higher than five for 85\% of Lyman-break galaxies and AGN \citep{garilli2021}.
Spectroscopic redshifts were determined for all objects using the Easy redshift (\texttt{EZ}) software package within the \texttt{PANDORA} environment \citep{garilli2010}.

        \begin{table*}[h]
        \caption[]{\texttt{slinefit} parameters.}
        \label{slinefit_params}
        \centering                          
        \begin{tabular}{l l l}        
        \hline\hline                 
        Parameter & Value & Description\\
        \hline                      
        dz                                      & 0.002 & Range of redshifts to explore around the input value\\
        delta\_z                                & 0.5   & Step in the grid of redshifts, given as the fraction of one pixel\\
        delta\_width                    & 0.5   & Step in the grid of line widths ($\sigma$), given as the fraction of one pixel\\
        delta\_offset                   & 0.5   & Step in the grid of line offsets, given as the fraction of one pixel\\
        num\_mc                                 & 200   & Number of random realisations of the spectrum to compute Monte Carlo uncertainties\\
        width\_min-width\_max   & 50-500        & Minimum and maximum allowed widths ($\sigma$) for a line [km/s]\\
        same\_width                             & false & Each line's width is allowed (true) to vary freely and independently in the fit, or not (false)\\
        offset\_max                     & $\pm$1000     & Maximum allowed velocity offset for lines [km/s]\\
        offset\_snr\_min                & 3     & Minimum S/N a line should have to be allowed to shift its centroid owing to velocity offsets\\
        \hline\hline
        \end{tabular}
        \end{table*}

        \begin{table}[h]
        \caption[]{Measured spectral lines (Gaussian fit).}
        \label{tabLines}
        \centering                          
        \begin{tabular}{|l c | l c|}        
        \hline                
        Ion$^{a}$ & Rest-frame         & Ion$^{a}$ & Rest-frame\\
                  & Wavelength$^{b}$ ($\AA$) &           & Wavelength$^{b}$ ($\AA$) \\
        \hline\hline                      
        CIII         &   1175.5      &          CII]         &   2326.0  \\
        Ly$\alpha$   &   1215.7      &          FeII         &   2344.2  \\
        NV           &   1240.8      &          FeII-1       &   2374.5  \\
        SiII         &   1260.4      &          FeII-2       &   2382.8  \\
        OI+SiII      &   1303.3      &          NeIV         &   2421.8  \\
        CII          &   1334.5      &          FeII-1       &   2586.7  \\
        OIV          &   1341.6      &          FeII-2       &   2600.2  \\
        SiIV-1       &   1393.8      &          MgII-1       &   2796.4  \\
        SiIV-2       &   1402.8      &          MgII-2       &   2803.5  \\
        NIV          &   1486.5      &          MgI          &   2853.0  \\
        SV           &   1501.8      &          NeV          &   3425.9  \\
        SiII         &   1526.7      &          OII          &   3727.4  \\
        CIV          &   1549.5      &          H$\beta$     &   4861.3  \\
        FeII         &   1608.5      &          OIII-1       &   4958.9  \\
        HeII         &   1640.4      &          OIII-2       &   5006.8  \\
        OIII]        &   1666.1      &          NII-2        &   6548.0  \\
        AlII         &   1670.8      &          H$\alpha$    &   6562.8  \\
        AlIII-1      &   1854.7      &          NII-1        &   6583.5  \\
        AlIII-2      &   1862.8      &          SII-1        &   6716.4  \\
        CIII]        &   1908.7      &          SII-2        &   6730.8  \\
        \hline
        \end{tabular}
        \begin{tablenotes}
        \small
        \item $^{a}$ Doublets are marked by the suffixes -1and -2.
        \item $^{b}$ Vacuum wavelengths are given for lines with $\lambda<3000\AA$; air wavelengths are given for lines with $\lambda>3000\AA$.
        \end{tablenotes}
        \end{table}

A redshift confidence flag was also assigned to each target, according to the following scheme, already applied to previous VIMOS surveys (e.g. VVDS, \citealp{lefevre2005}; zCOSMOS, \citealp{lilly2007}; VUDS, \citealp{lefevre2015}).
\begin{itemize}
        \item Flag 4: a highly reliable redshift (estimated to have a > 99\%
        probability of being correct), based on a high S/N spectrum
        and supported by obvious and consistent spectral features.
        \item Flag 3: also a very reliable redshift, comparable in confidence with Flag 4, 
        supported by clear spectral features in the
        spectrum, but not necessarily with a high S/N.
        \item Flag 2: a fairly reliable redshift measurement, although not as
        straightforward to confirm as those for Flags 3 and 4, supported by
        cross-correlation results, continuum shape, and some spectral
        features.
        \item Flag 1: a reasonable redshift measurement, based on weak
        spectral features and/or continuum shape.\\
        An \textit{\textit{\textit{\emph{a posteriori}}} }analysis of the redshift reliability showed that
        the reliability of Flag 2 redshifts is $\sim$79\%, 
        while that of Flag 1 redshifts is ~41\% \citep{garilli2021}.
        \item Flag 0: no reliable spectroscopic redshift measurement was
        possible.
        \item Flag 9: a redshift based on only one single clear spectral
        emission feature. An \textit{\emph{a posteriori}} analysis confirmed a redshift reliability
        of $\sim$ 95\% for spectra with this flag. 
        \item Flag -10: spectrum with clear problems in the observation or
        data-processing phases. 
        \item Flag 10+any of the above: broad line AGN (BLAGN). This preliminary
        classification has been subsequently revised by Bongiorno et al. (in preparation).
        \item Serendipitous (also called secondary) objects appearing by
        chance within the slit of the main target were identified by adding
        a '2' in front of the main flag.
\end{itemize}
The redshift accuracy, estimated by internal comparison between different observations, is $\sigma_{\Delta z/(1+z)}$ = 0.0007 \citep{garilli2021}.
The redshift distribution of the entire VANDELS sample is shown in Fig. \ref{zhist}.

In the official catalogues, we include only the measurements for the 1811 objects with a reliable redshift confidence flag (2, 3, 4, and 9 and the equivalent for BLAGN and secondary objects), whose redshift distribution is also shown in Fig. \ref{zhist}.

We measured spectroscopic features using two methods: Gaussian fit and direct integration. 

\section{Gaussian fit measurements}\label{sec:gauss}
Gaussian fit measurements were performed using \texttt{slinefit} \footnote{\url{https://github.com/cschreib/slinefit}} \citep{schreiber2018b}, an automated code that models the observed spectrum of a galaxy as a combination of a stellar continuum model and a set of emission and absorption lines. 

\subsection{\texttt{slinefit} parameters}
A set of templates from EAzY \citep{brammer2008}, based on the \citet{bruzual2003} stellar population models, is linearly combined to best fit the continuum. 
Table \ref{slinefit_params} summarises the parameters that were set to produce the official VANDELS catalogue.
The code searches for lines around their expected locations given by the input redshift:
lines with a S/N lower than \emph{offset\_snr\_min} are fixed at their expected position, while a velocity offset with respect to the measured redshift is allowed for lines with a higher S/N, with a maximum value set by the \emph{offset\_max} parameter.
We stress that in the catalogue the $\sigma$ of each line is provided, not the FWHM.

We measured 40 individual lines, including 7 resolved doublets, which are listed in Table \ref{tabLines}. 
Unresolved doublets (e.g. CIV$\lambda$1550 and CIII]$\lambda$1908) were treated as a single line. 
For the NII$\lambda\lambda$6548,6583 and SII$\lambda\lambda$6716,6730 doublets, we fixed the line flux ratios to 0.33:1 and 1:0.75, respectively, while no constraints were imposed for the other doublets. 
All lines were modelled as single symmetric Gaussians, either in emission or in absorption.  
This might not have been the best choice for the Ly$\alpha$ line, which typically is asymmetric and sometimes even split into a blue and a red component. 
Therefore, after visual inspection of the spectra by four members of the team, we added a flag indicating whether the fit was good (1) or not (0) and recommend using with caution the Ly$\alpha$ parameters from the catalogue in the latter cases.
In general, in the case of multi-component lines (e.g. P-Cygni profiles), only the strongest feature is fitted. 
We stress that \texttt{slinefit}, in our chosen configuration, always provides a solution. 
Therefore, we recommend caution when using spectral parameters when the lines are narrower than the spectral resolution (i.e. FWHM$\sim$460 km/s, corresponding to $\sigma\sim$195 km/s), because it might be a noise spike. 
Also, when the flux S/N$\leq$1, the line should be considered as undetected and the error on the flux can be used as a $1\sigma$ upper limit. 
Errors were evaluated through a Monte Carlo technique: the galaxy spectrum was randomly perturbed according to its re-scaled error spectrum (see the next Section) and the uncertainties on the spectroscopic parameters were then computed from the standard deviation of \emph{num\_mc} realisations of the fit.

        \begin{figure}[h]
        \centering
        \includegraphics[scale=0.43,trim=30mm 7mm 30mm 0mm, clip=true]{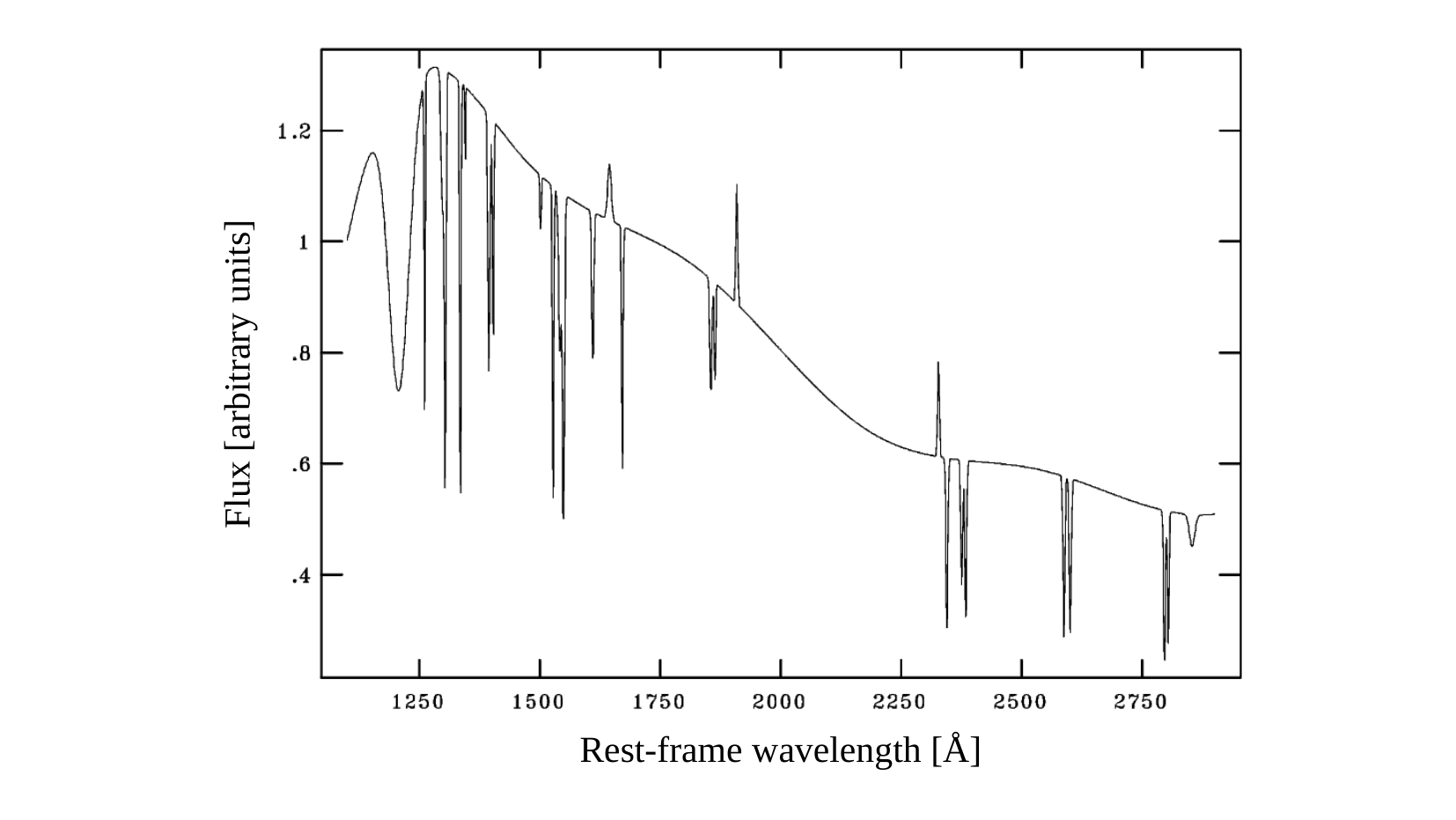}
        \includegraphics[scale=0.65,trim=65mm 10mm 75mm 15mm, clip=true]{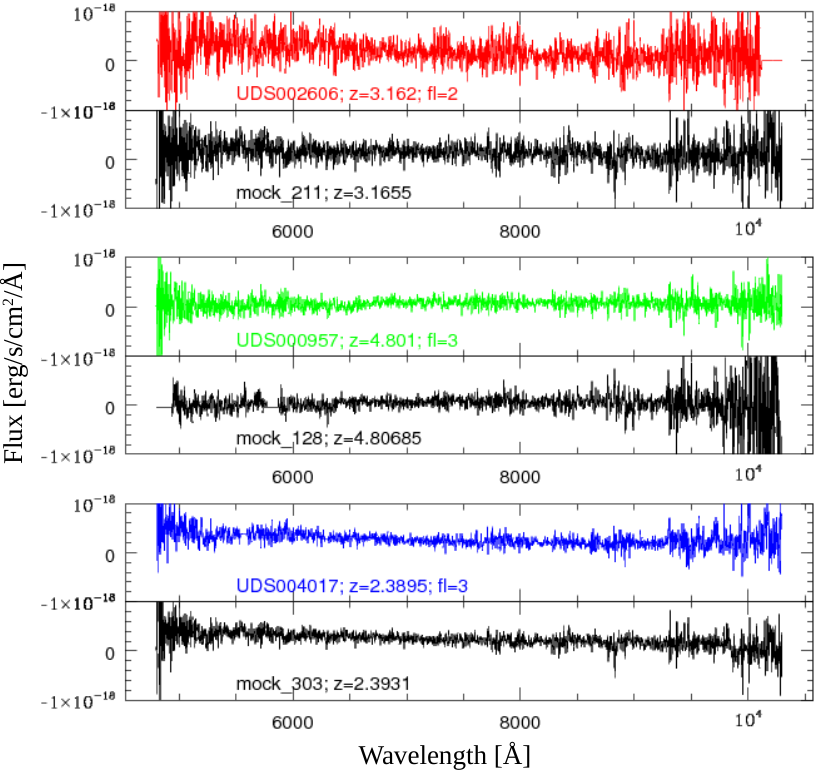}
        \caption{Construction of the 1D mock spectra for the \texttt{slinefit} code validation.
                \emph{Top}: Synthetic rest-frame template (from \citet{talia2012}), normalised to unity at 1750\AA.
                \emph{Bottom}: Comparison between three examples of mock 1D spectra and real VANDELS spectra. Mock spectra are shown in black. VANDELS spectra are colour-coded with respect to their depth: 20hrs (red), 40hrs (green), and 80hrs (blue). 
                }
        \label{rf_template}
        \end{figure}

        \begin{figure*}[h]
        \centering
        \includegraphics[scale=0.54,trim=50mm 1.5mm 60mm 0mm, clip=true]{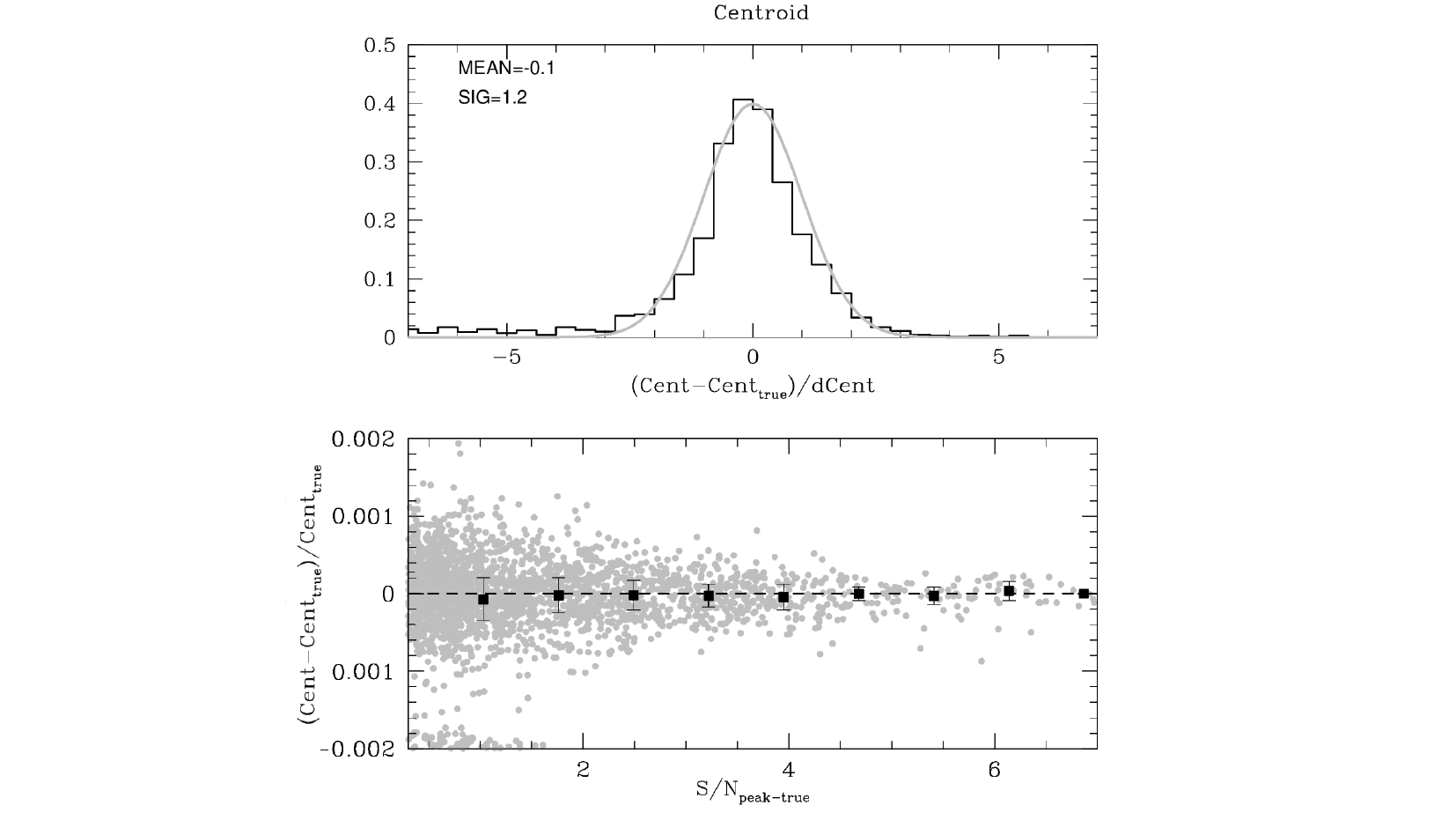}
        \includegraphics[scale=0.18,trim=0mm 5mm 0mm 0mm, clip=true]{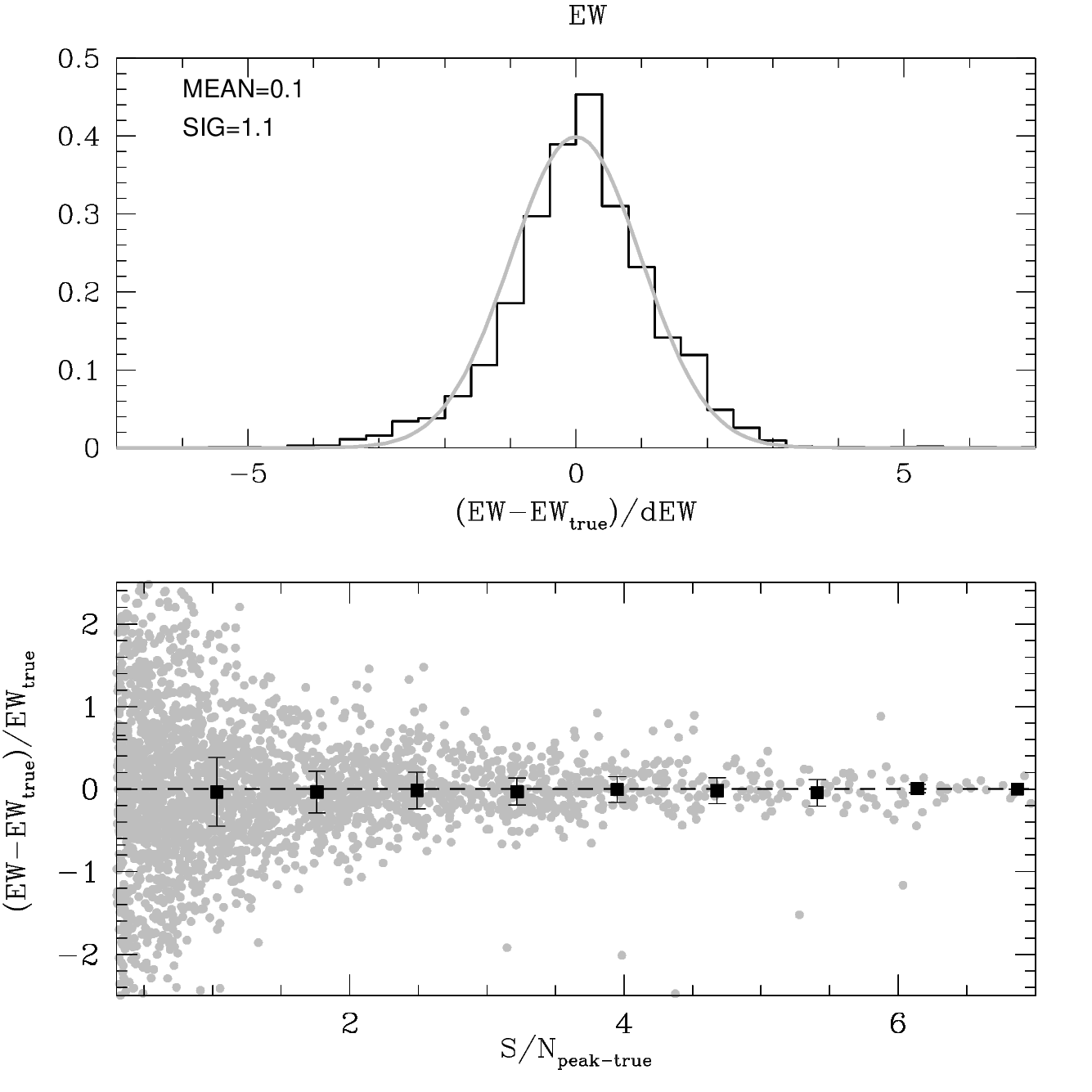}
        \includegraphics[scale=0.18,trim=0mm 5mm 0mm 0mm, clip=true]{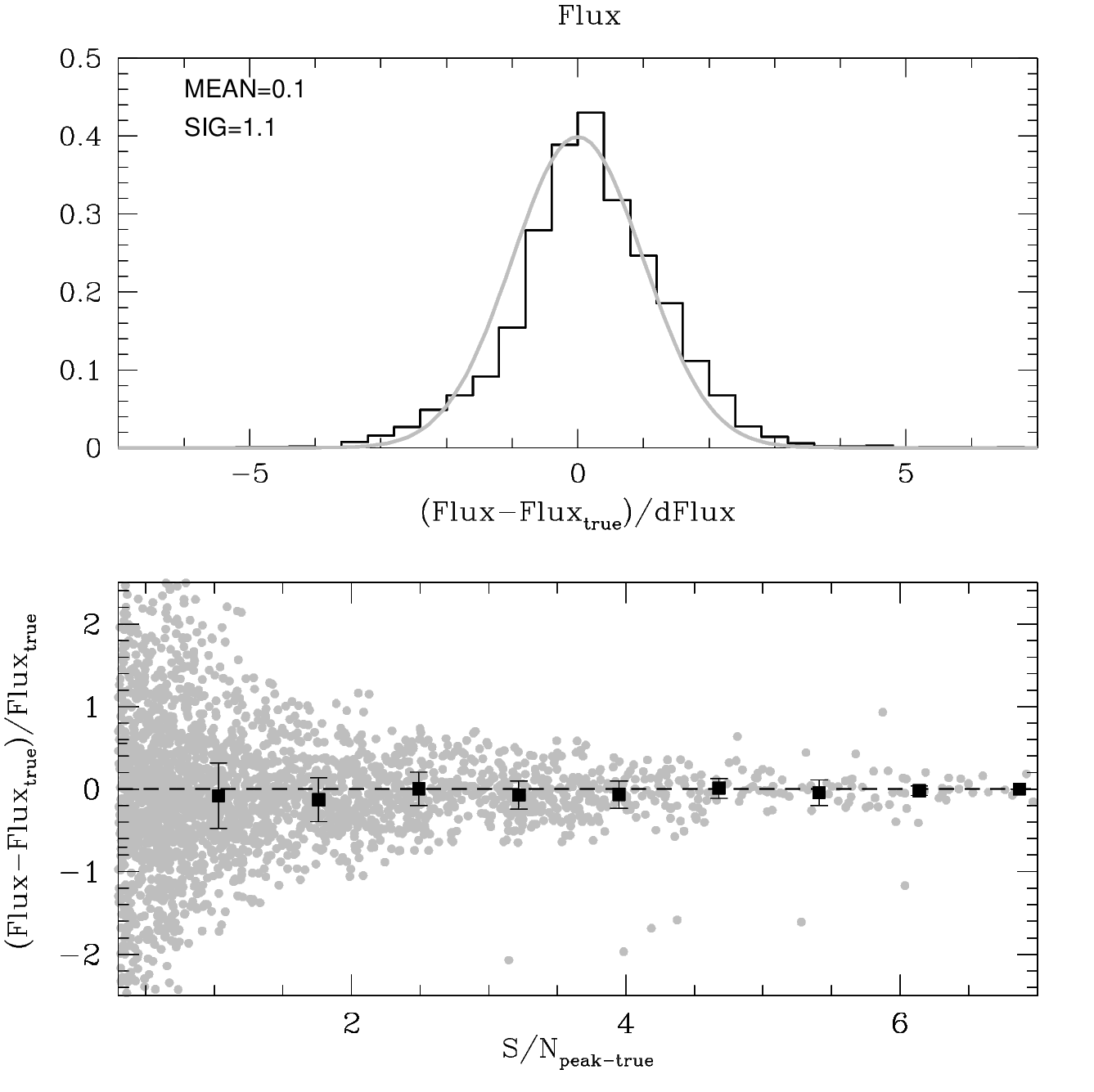}
        \includegraphics[scale=0.18,trim=0mm 5mm 0mm 0mm, clip=true]{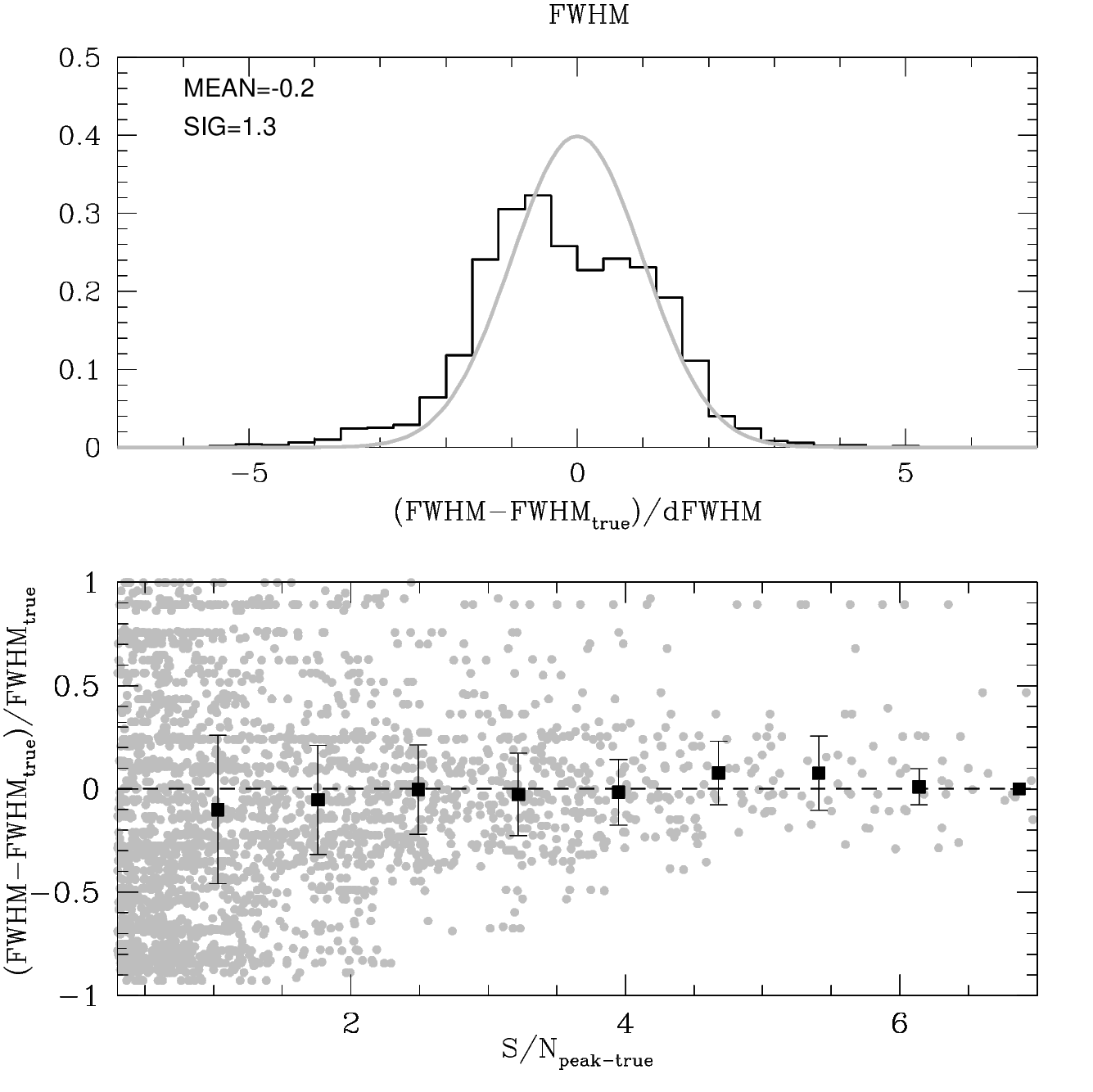}
        \caption{
                Comparison between \texttt{slinefit} results and input values for the sample of 270 mock spectra. 
                In the top panels, we plotted the pull distributions. As a reference, we marked with a grey curve a Gaussian with a null mean and unity sigma.
                In the bottom panels, we plotted the relative change of the measured spectral quantities, with respect to their input values, as a function of the peak S/N of the lines. Black squares represent the median values of the relative change in bins of the  S/N; error bars are the semi-interquartile range (SIQR).
                The line parameters are, starting clockwise from the top left figure: line centroid, EW, FWHM, and flux. 
                }
        \label{run_def}
        \end{figure*}

\subsection{Mock 1D spectra and \texttt{slinefit} code validation}\label{sec:mock}
In order to validate the \texttt{slinefit} code performance, we built a set of 1D mock spectra to mimic the characteristics of the observed VANDELS ones.
We started from a rest-frame template, normalised to unity at 1750\AA~ and created using the stacked spectrum of SFGs at z$\sim$2 from \citet{talia2012} as reference (Fig. \ref{rf_template}, top).
The continuum was modelled as a cubic spline, with a slope of $\beta \sim$-1.1 and a dispersion of 1\AA/pix. 
Emission and absorption spectral lines that are common in the UV range of SFG spectra (Table \ref{tabLines}) were added as symmetric Gaussians.
The lines were not all added  at their vacuum rest-frame wavelength: some shifts were introduced in order to mimic the effects of outflows.
We created three templates with the same continuum and varying the lines' peak S/N in the range 0.3--7.
Then, each rest-frame template was used to create 30 redshifted templates, with redshifts evenly distributed in the range 2.2--5.

The redshifted templates were normalised to the F814W observed magnitude, following the magnitude versus redshift relation of the VANDELS survey.
They were then re-sampled and cut to the VANDELS dispersion and observed wavelength range.

In order to add realistic noise, we extracted 1D spectra from empty regions in observed 2D spectra from the VANDELS survey at different exposure times and added them to the redshifted templates.
The final validation sample counts 270 mock spectra. 
In Fig. \ref{rf_template} (bottom) we show the comparison between three examples of mock 1D spectra and real VANDELS spectra at different redshifts and with different quality flags.

Finally, we ran the \texttt{slinefit} on the sample of mock 1D spectra with different sets of input parameters and checked the relative change in the measured spectral quantities with respect to their input values, and the pull distributions.
In Fig. \ref{run_def} we show the results from the run with the best set of parameters, which is summarised in Table \ref{slinefit_params}.
All the measured lines are included in the plots, but we stress that separating emission and absorption lines does not change the results.
All distributions are consistent with a Gaussian with a null mean and unity sigma.

        \begin{figure}
        \centering
        \includegraphics[scale=0.55,trim=60mm 50mm 60mm 50mm, clip=true]{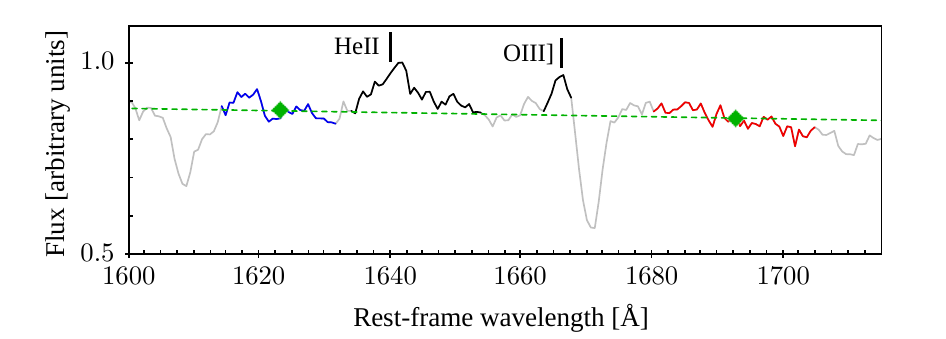}
        \includegraphics[scale=0.55,trim=60mm 50mm 60mm 50mm, clip=true]{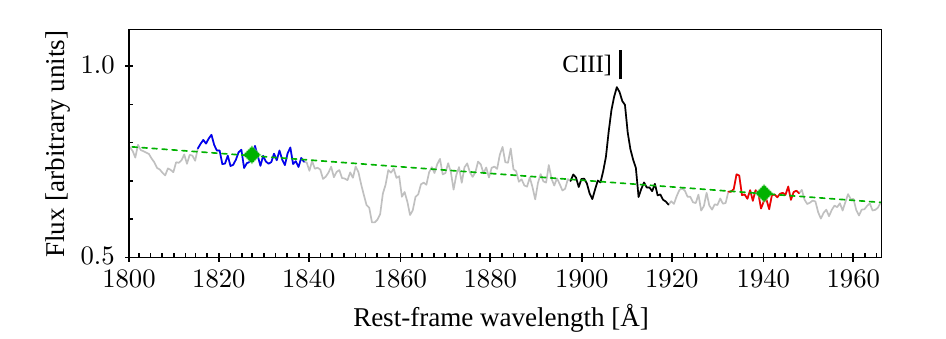}
        \caption{Median composite spectrum of VANDELS sources (grey). The upper and lower panels show zoomed-in regions around the HeII$\lambda$1640+OIII]$\lambda$1666 and CIII]$\lambda$1909 lines, respectively. The central bandpasses, as indicated in Table \ref{tabIndex}, are marked in black, while the two local continuum windows are marked in blue and red. The green points and dashed lines indicate the mean flux in the continuum bandpasses and the linear pseudo-continuum. 
        }
        \label{dr2_stack}
        \end{figure}

\section{Direct integration}\label{sec:lick}
The direct integration measurements were performed using \texttt{pylick} \footnote{\url{https://gitlab.com/mmoresco/pylick/}}, a flexible Python tool to measure spectral indices and associated uncertainties. 
The code is described in \citet{borghi2021} and was extensively tested using spectra and results from the LEGA-C survey \citep{vanderwel2016, straatman2018}. 
Following the approach of the Lick group \citep{worthey1997}, the code computes the strengths of a set of atomic and molecular indices and continuum breaks such as the D4000 \citep{bruzual1983}. 
Errors are evaluated following the S/N method by \citet{cardiel1998}.

        \begin{figure}
        \centering
        \includegraphics[scale=0.24]{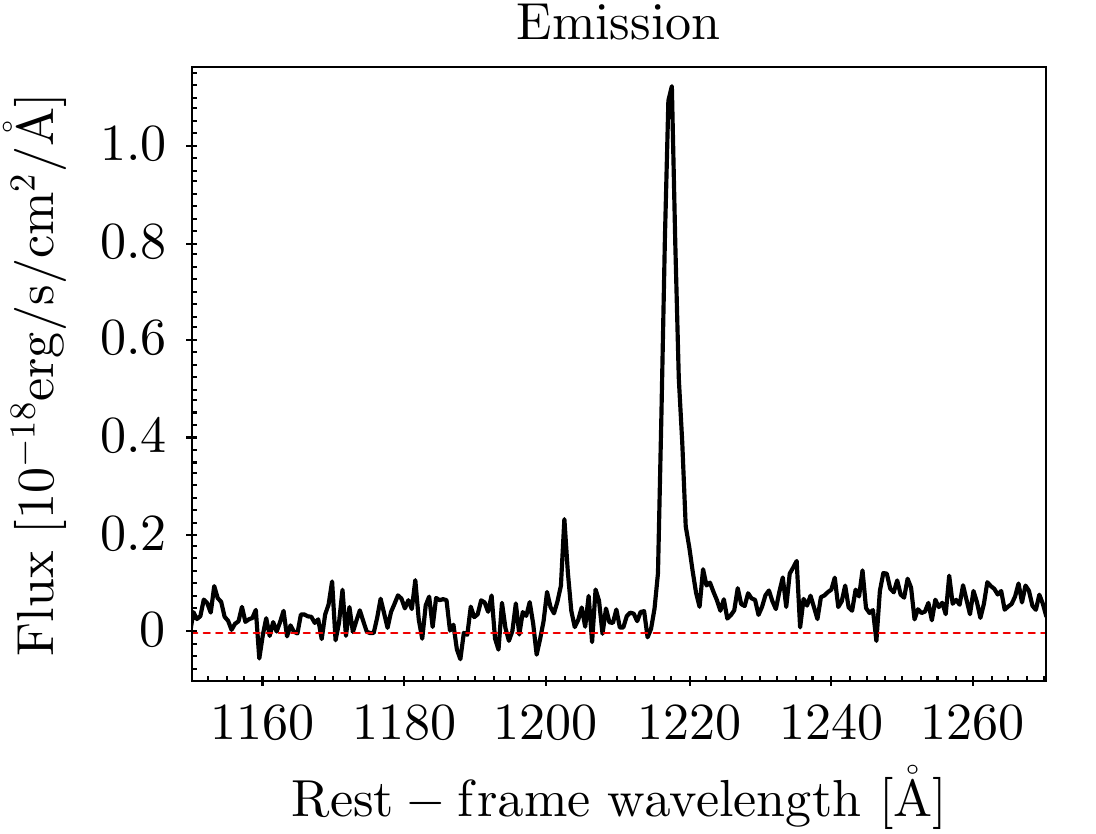}\includegraphics[scale=0.24]{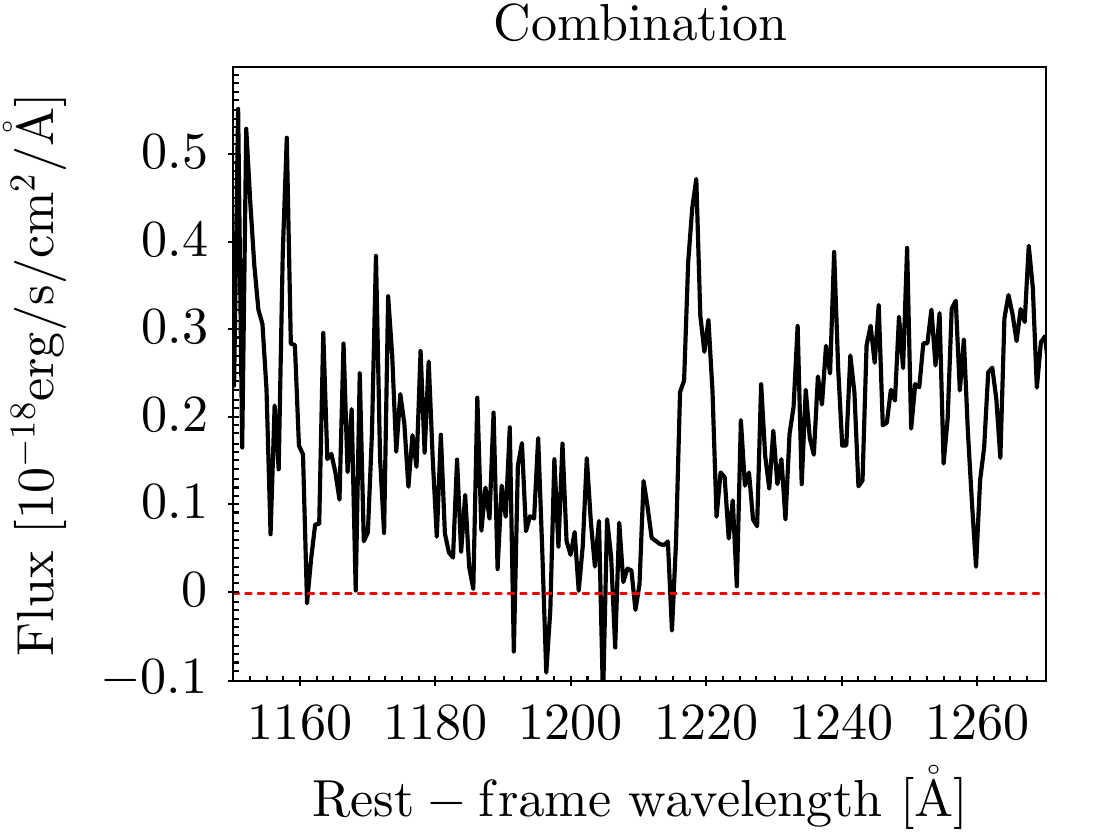}
        \includegraphics[scale=0.24]{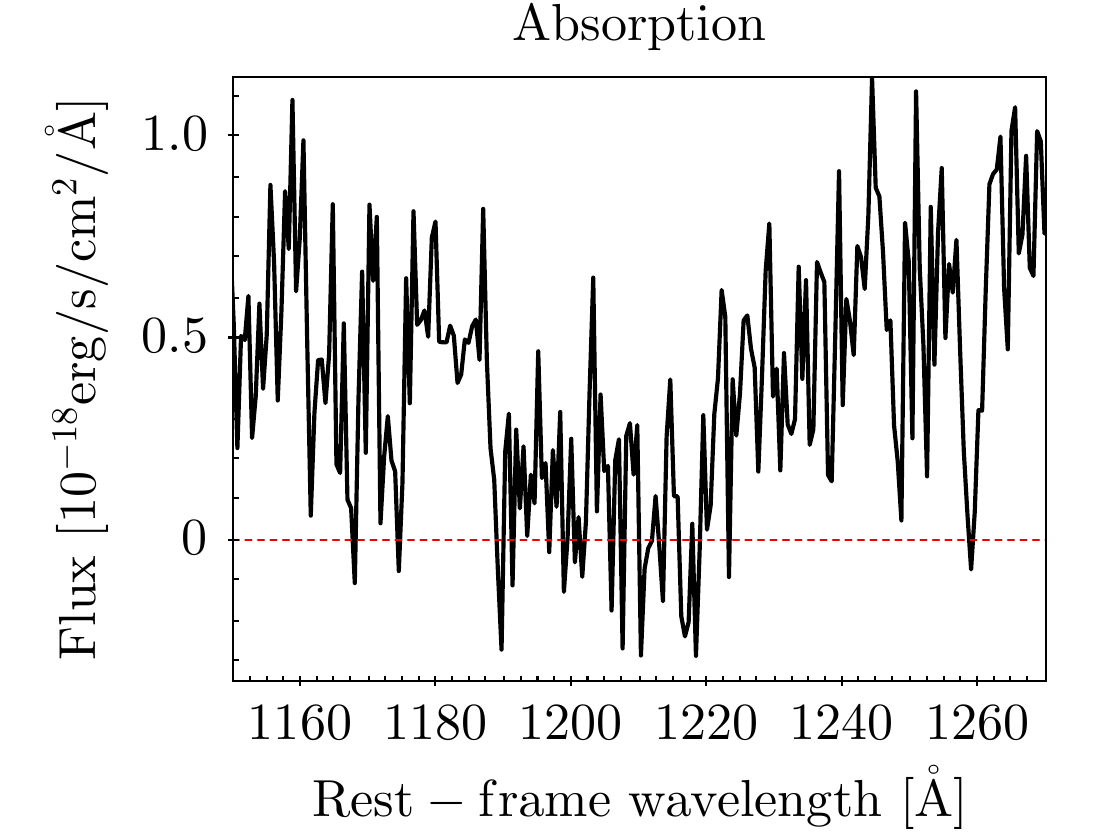}\includegraphics[scale=0.24]{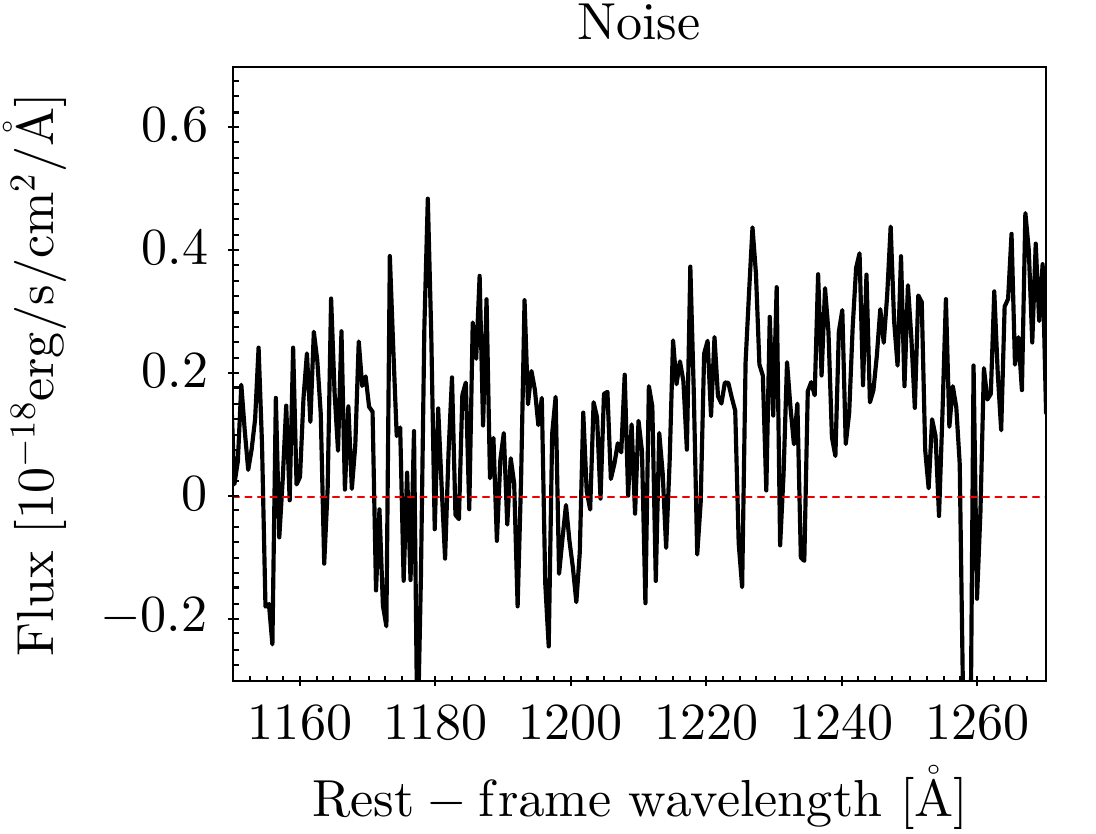}
        \caption{Examples illustrating the four Ly$\alpha$ categories from \citet{kornei2010}. Clockwise, from the top left: emission, combination, absorption, and noise. 
        } 
        \label{lya_class}
        \end{figure}

In our catalogue, we have included 55 
indices and breaks defined in previous works, to which we added three UV emission line indices (see Table \ref{tabIndex}).
The new indices were defined on the basis of a high-S/N composite spectrum of all VANDELS sources with a high-redshift quality flag (i.e. 3 and 4). 
It was built by median stacking the de-redshifted, scaled (by the median flux in the wavelength range 1410--1510~\AA), and rebinned (0.6~\AA/pix) spectra. 
In Fig. \ref{dr2_stack} we show the zoomed-in regions around the HeII$\lambda$1640+OIII]$\lambda$1666 and CIII]$\lambda$1909 lines, with the central bandpass and pseudo-continuum ranges marked in different colours.
It should be noted that for the direct integration catalogue, no offset of the bandpasses is allowed with respect to the expected wavelength, given the redshift. 

For the Ly$\alpha$ line, we opted for a different approach. 
Following \citet{cullen2020}, we applied the method by \citet{kornei2010} to measure the EW of the line, which takes into account the line's morphology to optimise the wavelength range over which the flux is integrated. 
The Ly$\alpha$ line of the 1218 individual galaxies at z$\gtrsim$2.95 (i.e. the redshift limit for the Ly$\alpha$ to be in the VIMOS wavelength range) was visually classified as either 
\emph{emission}, \emph{absorption}, \emph{combination}, or \emph{noise}; examples are shown in Fig. \ref{lya_class}.
The \emph{emission} spectra are clearly dominated by a Ly$\alpha$ emission feature. 
The \emph{absorption} spectra are dominated by an extended trough around the Ly$\alpha$ position. 
In the \emph{combination} case, the spectrum contain superimposed emission and absorption features.
Finally, the \emph{noise} category include spectra where no clear feature could be identified at the Ly$\alpha$ position \citep[see][for a detailed description]{kornei2010}.
In the first three cases, after the peak of the emission and absorption, the integration window is defined by the wavelength values on either side of the peak where the flux intersects the average continuum level. 
The blue and red continua are defined as the median flux values in the range $\lambda$=[1120--1180]\AA~ and [1228--1255]\AA, respectively. 
In the case of  \emph{absorption} and \emph{combination} sources, the spectra were first smoothed with a boxcar function of  six pixels in width to minimise the possibility of noise spikes affecting the determination of the boundaries of the integration range. 
For \emph{noise} sources, the Ly$\alpha$ flux is simply defined as the integrated flux in the range $\lambda$=[1200--1228]\AA. 
In all cases the line flux was divided by the red continuum value to obtain the EW.

\section{Scaling of the error spectra}\label{sec:errspec}
\begin{figure}
        \centering
        \includegraphics[scale=0.5]{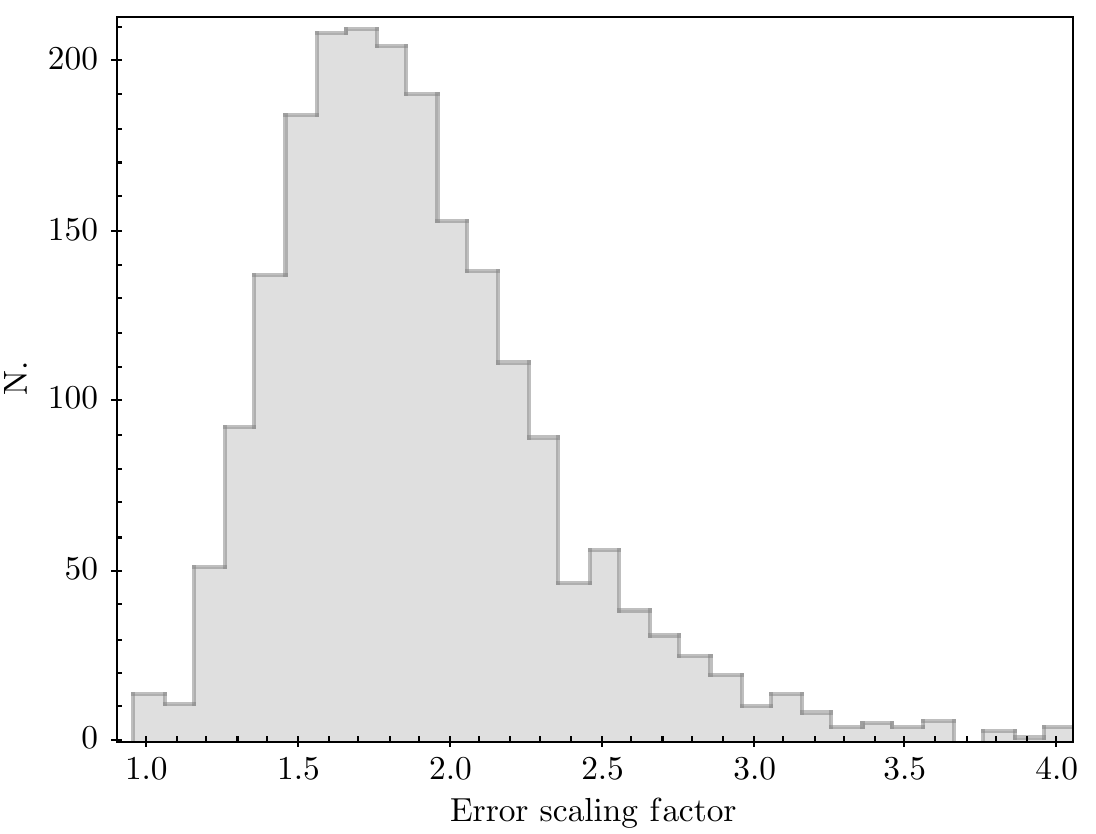}
        \caption{Distribution of the multiplicative scaling factors to correct the mismatch between the error spectra and the noise of the object spectra. 
        } 
        \label{scale_fac}
        \end{figure}

The spectra distributed as part of the VANDELS public data release include the 1D noise estimate\footnote{This is the extension NOISE in the multi-extension FITS files distributed through the VANDELS collaboration website and the column ERR in the FITS binary tables downloadable from the ESO archive.} in erg cm$^{-2}$ s$^{-1}$ \AA$^{-1}$.
The \emph{error} spectrum is a direct product of the data reduction procedures \citep{garilli2021}, and should reflect the noise level of the corresponding object spectrum.
However, the comparison between the error spectra and the noise r.m.s. of the object spectra, measured in line-free regions, shows a discrepancy, with the error spectra underestimating the noise level by a factor of $\sim$2, on average.
We performed several tests on 2D and 1D spectra: our hypothesis is that the discrepancy is caused by the fact that the data reduction pipeline does not take the full covariance matrix into account.
We opted for an \emph{\emph{a posteriori}} statistical correction of the error spectra (e.g. \citet{vanderwel2021}). 
In particular, for each object, we computed a scaling factor to be applied to the error spectrum. 
The scaling factor is defined as the standard deviation of the fit residuals, divided by the error spectrum:
\begin{align*}
1.482\times MAD[(object_{spectrum} - model_{spectrum}) / error_{spectrum}],
\end{align*}
where model$_{spectrum}$ is the output of \texttt{slinefit} and MAD is the median absolute deviation.
If the error spectrum is an accurate representation of the noise in the object spectrum, the above quantity should be close to 1; if the error spectrum underestimates the noise, then the above quantity can be used as a scaling factor.
We computed it in five wavelength windows, free of strong sky lines, and then defined the scaling factor as the mean of the five values.
The associated uncertainty is the error on the mean, which takes into account a slight wavelength dependence of the ratio between the noise r.m.s. of the object spectrum and the error spectrum (i.e. the ratio is on average $\sim$10$\%$ lower close to the spectral edges than in the central region).
Fig. \ref{scale_fac} shows the distribution of the scaling factor.

The \texttt{slinefit} code can actually perform the scaling of the error spectrum internally.
If the appropriate keyword (\texttt{residual\_{rescale}}) is switched on, the previously defined scaling factor is computed locally for each line; then, the whole error spectrum is normalised by interpolating between the scaling factors of the chosen lines, and the whole fit is performed a second time.
The measurements in the official catalogue were instead performed by applying a single scaling factor to each error spectrum before running \texttt{slinefit} with the \texttt{residual\_{rescale}} keyword switched off.
This choice allowed us to provide a set of measurements that could be easily reproduced by other codes that do not include a scaling feature.
The direct integration measurements  were also performed \emph{after} scaling the error spectra.

The scaling factors and their uncertainties are included in both catalogues.
We stress that the error spectra in the VANDELS data release (i.e. NOISE extension or ERR column) are {\bf not} scaled: they have to be multiplied by the scaling factor in order to obtain reliable errors on the spectroscopic measurements.

        \begin{table*}
        \caption[]{\texttt{pylick} spectral indices.}
        \label{tabIndex}    
        \small 
        \centering                          
        \begin{tabular}{l c c c c c c}        
        \hline\hline                 
        Index  & Central Bandpass & Blue Continuum & Red Continuum & Units & Type & Ref.$^{a}$\\
        \hline    
        \hline              
        BL$_{1302}$  &  1292.000--1312.000 &   1270.000--1290.000  &  1345.000--1365.000  &       \AA     & atomic$^{b}$                          &  3   \\
        OI+SiII1303  &  1290.000--1307.000 &   1268.000--1286.000  & 1308.000--1324.000  &       \AA     & atomic                                        &  9   \\
        CII1335         & 1326.000--1340.000    & 1308.000--1324.000    & 1348.000--1378.000  &    \AA     & atomic                                        &  9    \\
        SiIV1400        & 1380.000--1407.000    & 1348.000--1378.000    & 1433.000--1460.000  &    \AA     & atomic                                        &  this work$^{f}$      \\
        SiIV         &  1387.000--1407.000 &   1345.000--1365.000  &  1475.000--1495.000  &       \AA     & atomic                                        &  3   \\
        BL$_{1425}$  &  1415.000--1435.000 &   1345.000--1365.000  &  1475.000--1495.000  &       \AA     & atomic                                        &  3   \\
        Fe1453       &  1440.000--1466.000 &   1345.000--1365.000  &  1475.000--1495.000  &       \AA     & atomic                                        &  3   \\
        SiII1526        & 1521.000--1533.000    & 1460.000--1495.000    & 1572.000--1599.000  &    \AA     & atomic                                        &  this work$^{f}$      \\
        C$_{IV}^{A}$ &  1530.000--1550.000 &   1500.000--1520.000  &  1577.000--1597.000  &          \AA  & atomic                                        &  3   \\
        CIV              &  1540.000--1560.000 &   1500.000--1520.000  &  1577.000--1597.000  &   \AA     & atomic                                        &  3    \\
        C$_{IV}^{E}$ &  1550.000--1570.000 &   1500.000--1520.000  &  1577.000--1597.000  &          \AA  & atomic                                        &  3   \\
        FeII1608        & 1600.000--1616.000    & 1583.000--1599.000    & 1614.000--1632.000  &    \AA     & atomic                                        &  this work$^{f}$      \\
        BL$_{1617}$  &  1604.000--1630.000 &   1577.000--1597.000  &  1685.000--1705.000  &       \AA     & atomic                                        &  3   \\
        HeII 1640       & 1634.000--1654.000    & 1614.000--1632.000    & 1680.000--1705.000  &    \AA     & atomic                                        &  this work    \\
        BL$_{1664}$  &  1651.000--1677.000 &   1577.000--1597.000  &  1685.000--1705.000  &       \AA     & atomic                                        &  3   \\
        OIII] 1666      & 1663.000--1668.000    & 1614.000--1632.000    & 1680.000--1705.000  &    \AA     & atomic                                        &  this work    \\
        AlII1670        & 1663.000--1679.000    & 1614.000--1632.000    & 1680.000--1705.000  &    \AA     & atomic                                        &  9    \\
        BL$_{1719}$  &  1709.000--1729.000 &   1685.000--1705.000  &  1803.000--1823.000  &       \AA     & atomic                                        &  3   \\
        BL$_{1853}$  &  1838.000--1868.000 &   1803.000--1823.000  &  1885.000--1915.000  &       \AA     & atomic                                        &  3   \\
        AlIII1860       & 1840.000--1873.000    & 1815.000--1839.000    & 1932.000--1948.000  &    \AA     & atomic                                        &  9    \\
        CIII 1909       & 1897.000--1919.000    & 1815.000--1839.000    & 1932.000--1948.000  &    \AA     & atomic                                        &  this work    \\
        FeII2370     &  2334.000--2391.000 &   2267.000--2290.000  &  2395.000--2450.000  &      \AA     & atomic				 &  9 \\
        FeII2402     &  2382.000--2422.000 &   2285.000--2325.000  &  2432.000--2458.000  &       \AA     & atomic                                        &  3   \\
        BL$_{2538}$  &  2520.000--2556.000 &   2432.000--2458.000  &  2562.000--2588.000  &       \AA     & atomic                                        &  3   \\
        FeII2600     &  2578.000--2611.000 &   2525.000--2572.000  &  2613.000--2674.000  &      \AA     & atomic     				 &  9 \\
        FeII2609     &  2596.000--2622.000 &   2562.000--2588.000  &  2647.000--2673.000  &       \AA     & atomic                                        &  3   \\
        B(2640)      &                     &   2600.000--2630.000  &  2645.000--2675.000  &       dex     & break$_{\lambda}$$^{c}$   &  7       \\
        MgII2800     &  2788.000--2810.000 &   2720.000--2785.000  &  2812.000--2842.000  &      \AA     & atomic				 &  9 \\
        MgII         &  2784.000--2814.000 &   2762.000--2782.000  &  2818.000--2838.000  &       \AA     & atomic                                        &  3   \\
        MgI          &  2839.000--2865.000 &   2818.000--2838.000  &  2906.000--2936.000  &   \AA     & atomic                                        &  3    \\
        Mg$_{UV}$    &  2625.000--2725.000 &   2525.000--2625.000  &  2725.000--2825.000  &       dex     & bump$^{d}$                            &  4   \\
        Mg$_{wide}$  &  2670.000--2870.000 &   2470.000--2670.000  &  2930.000--3130.000  &       \AA     & atomic                                        &  3   \\
        B(2900)      &                     &   2855.000--2885.000  &  2915.000--2945.000  &       dex     & break$_{\lambda}$$^{c}$   &  7       \\
        FeI              &  2965.000--3025.000 &   2906.000--2936.000  &  3031.000--3051.000  &   \AA     & atomic                                        &  3    \\
        BL$_{3096}$  &  3086.000--3106.000 &   3031.000--3051.000  &  3115.000--3155.000  &       \AA     & atomic                                        &  3   \\
        CaII~K       &  3925.650--3945.000 &   3845.000--3880.000  &  3950.000--3954.000  &       \AA     & atomic                                        &  8   \\
        CaII~H       &  3959.400--3975.000 &   3950.000--3954.000  &  3983.000--3993.000  &       \AA     & atomic                                        &  8   \\
        D4000        &                     &   3750.000--3950.000  &  4050.000--4250.000  &       dex     & break$_{\nu}$                         &  5   \\
        D$_{n}$4000  &                     &   3850.000--3950.000  &  4000.000--4100.000  &       dex     & break$_{\nu}$                         &  6   \\
        H$\delta_{A}$&  4083.500--4122.250 &   4041.600--4079.750  &  4128.500--4161.000  &       \AA     & atomic                                        &  2   \\
        H$\delta_{F}$&  4091.000--4112.250 &   4057.250--4088.500  &  4114.750--4137.250  &       \AA     & atomic                                        &  2   \\
        CN$_1$       &  4142.125--4177.125 &   4080.125--4117.625  &  4244.125--4284.125  &       mag     & molecular$^{e}$                       &  1   \\
        CN$_2$       &  4142.125--4177.125 &   4083.875--4096.375  &  4244.125--4284.125  &       mag     & molecular                             &  1   \\
        Ca4227       &  4222.250--4234.750 &   4211.000--4219.750  &  4241.000--4251.000  &       \AA     & atomic                                        &  1   \\
        G4300        &  4281.375--4316.375 &   4266.375--4282.625  &  4318.875--4335.125  &       \AA     & atomic                                        &  1   \\
        H$\gamma_{A}$&  4319.750--4363.500 &   4283.500--4319.750  &  4367.250--4419.750  &       \AA     & atomic                                        &  2   \\
        H$\gamma_{F}$&  4331.250--4352.250 &   4283.500--4319.750  &  4354.750--4384.750  &       \AA     & atomic                                        &  2   \\
        Fe4383       &  4369.125--4420.375 &   4359.125--4370.375  &  4442.875--4455.375  &       \AA     & atomic                                        &  1   \\
        Ca4455       &  4452.125--4474.625 &   4445.875--4454.625  &  4477.125--4492.125  &       \AA     & atomic                                        &  1   \\
        Fe4531       &  4514.250--4559.250 &   4504.250--4514.250  &  4560.500--4579.250  &       \AA     & atomic                                        &  1   \\
        C$_{2}$4668  &  4634.000--4720.250 &   4611.500--4630.250  &  4742.750--4756.500  &       \AA     & atomic                                        &  1   \\
        H$\beta$     &  4847.875--4876.625 &   4827.875--4847.875  &  4876.625--4891.625  &       \AA     & atomic                                        &  1   \\
        Fe5015       &  4977.750--5054.000 &   4946.500--4977.750  &  5054.000--5065.250  &       \AA     & atomic                                        &  1   \\
        Mg$_1$       &  5069.125--5134.125 &   4895.125--4957.625  &  5301.125--5366.125  &       mag     & molecular                             &  1   \\
        Mg$_2$       &  5154.125--5196.625 &   4895.125--4957.625  &  5301.125--5366.125  &       mag     & molecular                             &  1   \\
        Mg$_b$       &  5160.125--5192.625 &   5142.625--5161.375  &  5191.375--5206.375  &       \AA     & atomic                                        &  1   \\
        Fe5270       &  5245.650--5285.650 &   5233.150--5248.150  &  5285.650--5318.150  &       \AA     & atomic                                        &  1   \\
        Fe5335       &  5312.125--5352.125 &   5304.625--5315.875  &  5353.375--5363.375  &       \AA     & atomic                                        &  1   \\
        \hline\hline
        \end{tabular}
        \small
        \begin{tablenotes}
        \item $^{a}$ 1: \citet{trager1998}; 2: \citet{worthey1997}; 3: \citet{maraston2009}; 4: \citet{daddi2005}; 5: \citet{bruzual1983}; 6: \citet{balogh1999}; 7: \citet{spinrad1997}; 8: \citet{fanfani2019}; 9: \citet{leitherer2011}. 
        \item $^{b}$ \citet{borghi2021} Eq. 1; $^{c}$ \citet{borghi2021} Eq. 2; $^{d}$ \citet{borghi2021} Eq. 4; $^{e}$ as \citet{borghi2021} Eq. 4, but integrating over $F(\lambda)d\lambda$, instead of $F(\nu)d\nu$; $^{f}$ these indices were firstly defined by \citet{leitherer2011}, but here we present a slightly modified version.
        \end{tablenotes}
        \end{table*}

        \begin{table*}
        \caption[]{Legend of catalogue content.}
        \label{col}
        \centering                          
        \begin{tabular}{l l l l}        
        \hline\hline                
        \# & Column name & Description & Units \\
        \hline
        \multicolumn{4}{c} {Gaussian fit} \\ 
        \hline         
        1 & objID & Object identification &\\
        2 & z & Spectroscopic redshift &\\
        3 & zQfl & Redshift confidence flag (Sec. \ref{sec:vandels}) &\\
        4-5 & scaling\_factor, e\_scaling\_factor &Scaling factor (Sec. \ref{sec:errspec}) (and error) &\\
        6 & ly$\alpha$\_fl & Goodness-of-fit flag for Ly$\alpha$ (Sec. \ref{sec:gauss}) &\\
        7-406 &  & Lines parameters from \texttt{slinefit}: &\\
        & wave, ewave & observed centroid wavelength (and error)& microns\\
        & flux, eflux & lines flux (and error)& erg s$^{-1}$ cm$^{-2}$\\ 
        & cont, econt & continuum flux (and error)&  erg s$^{-1}$ cm$^{-2}$ \AA$^{-1}$\\
        & ew, eew & rest-frame EW$^{a}$ (and error)& \AA \\
        & sigma, esigma & line width$^{b}$ (and error)& km s$^{-1}$\\
        \hline
        \multicolumn{4}{c} {Direct integration} \\  
        \hline          
        1 & objID & Object identification &\\
        2 & z & Spectroscopic redshift &\\
        3 & zQfl & Redshift confidence flag (Sec. \ref{sec:vandels}) &\\
        4-5 & err\_scaling\_factor, eerr\_scaling\_factor &Error scaling factor (Sec. \ref{sec:errspec}) (and error) &\\
        6 & lya\_class & Ly$\alpha$ visual classification (Sec. \ref{sec:lick}): &\\
        &&1=emission; 2=combination; 3=absorption; 4=noise &\\
        7-8 & LyA\_EW0\_K10, LyA\_EW0\_K10\_err & Ly$\alpha$ EW following \citet{kornei2010} (and error) & \AA \\
        &  & Indices/breaks from \texttt{pylick}: &\\        
        9-182 & \emph{index name}, err & rest-frame EW$^{a}$ [atomic indices]; & \AA; \\
        &&rest-frame EW [molecular indices]; break (and error) & mag; unitless (Tab. \ref{tabIndex})\\
        & cont & Pseudo-continuum flux & erg s$^{-1}$ cm$^{-2}$ \AA$^{-1}$\\
        \hline\hline
        \end{tabular}
        \small
        \begin{tablenotes}
        \item $^{a}$ EW sign convention: positive for emission lines; negative for absorption lines.
        \item $^{b}$ We stress that in the catalogue the Gaussian $\sigma$ of each line is provided, not the FWHM.
        \end{tablenotes}
        \end{table*}

\section{The catalogues}\label{sec:cat}
\begin{figure*}
        \centering
        \includegraphics[scale=0.47]{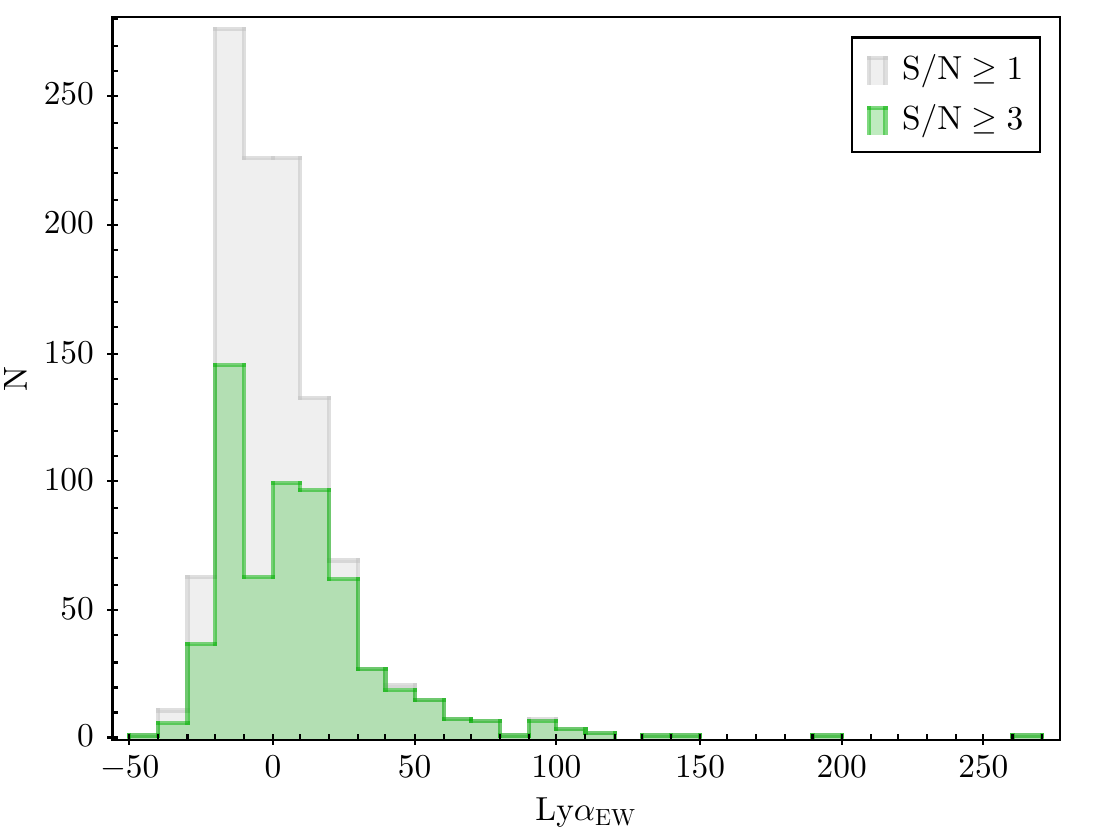}
        \includegraphics[scale=0.47]{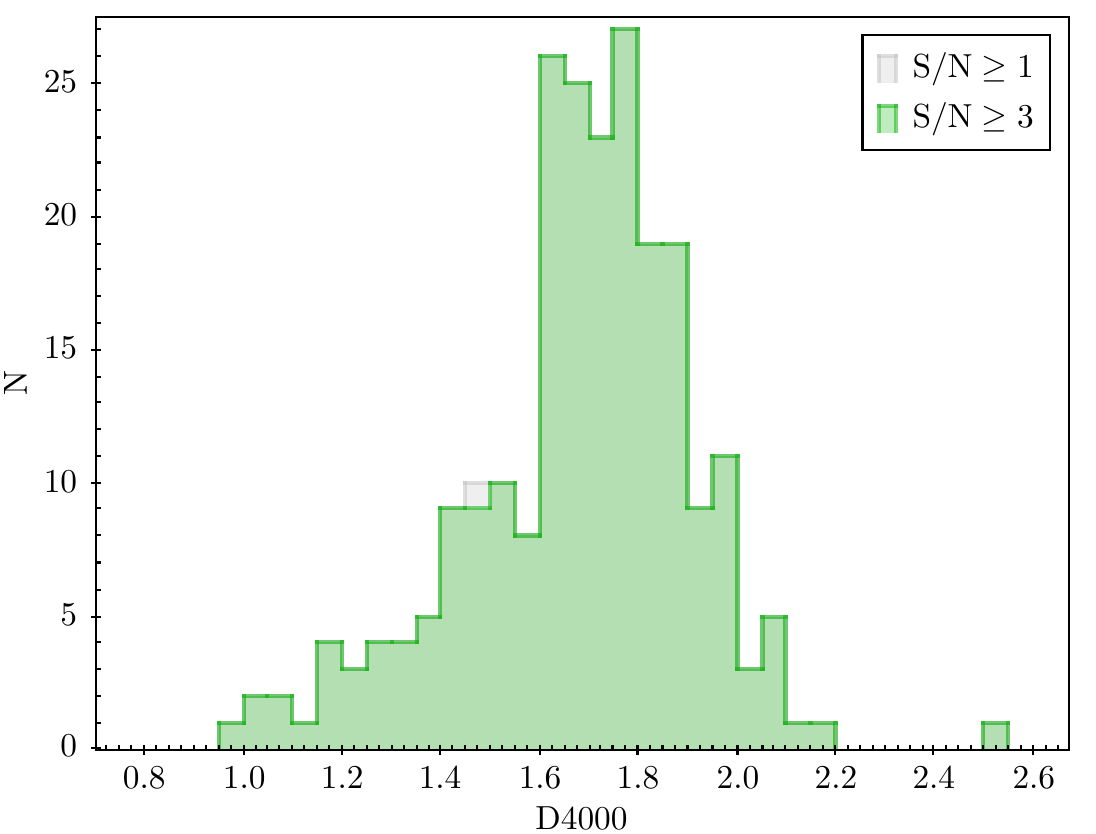}
        \includegraphics[scale=0.47]{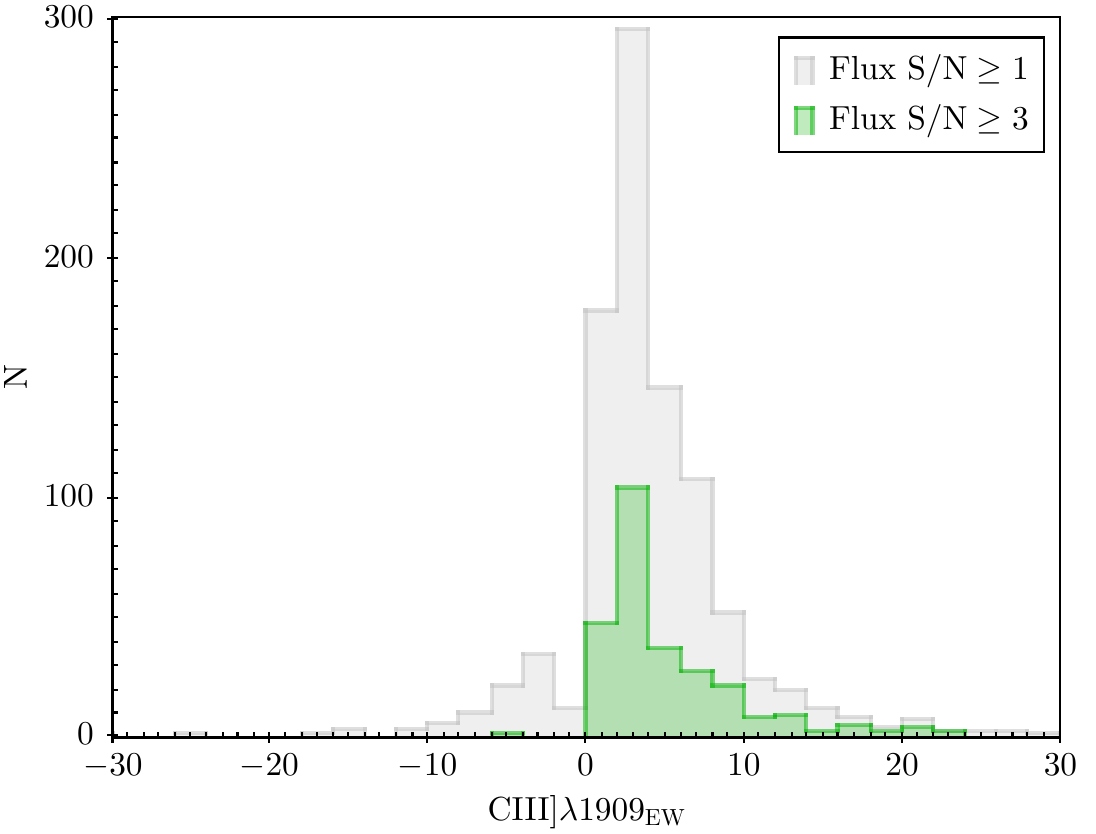}
        \includegraphics[scale=0.47]{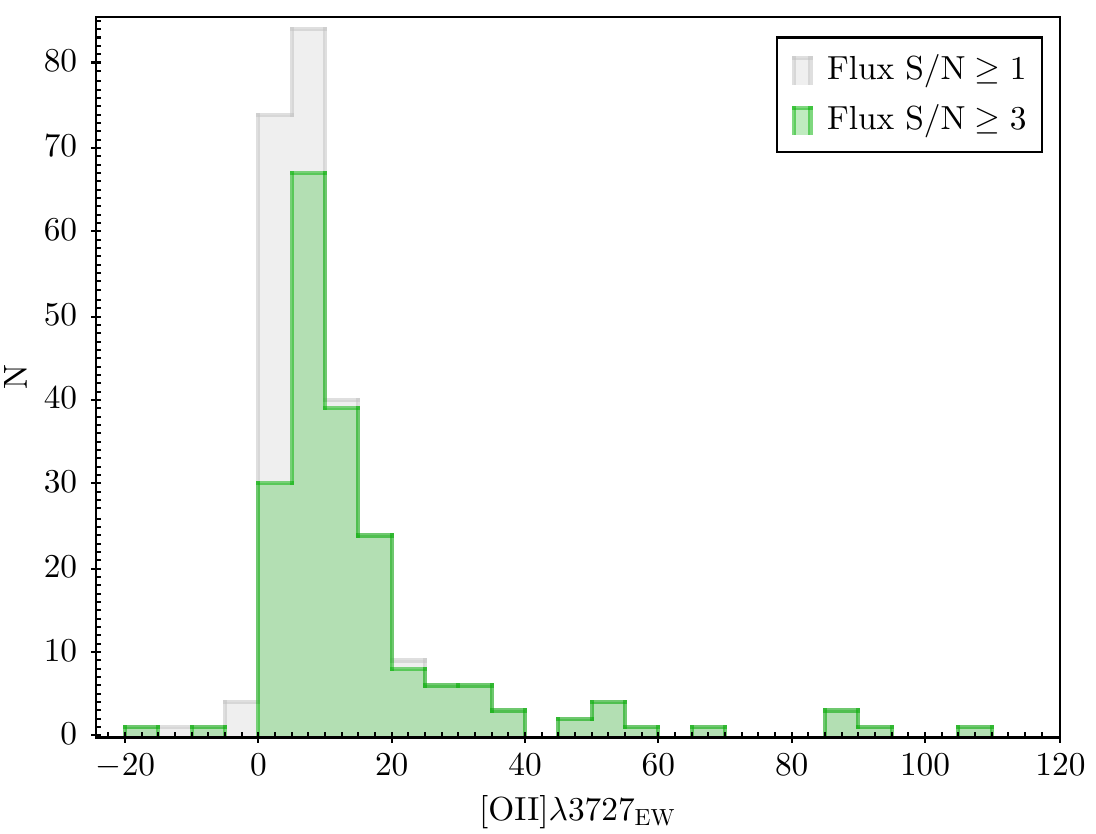}
        \includegraphics[scale=0.47]{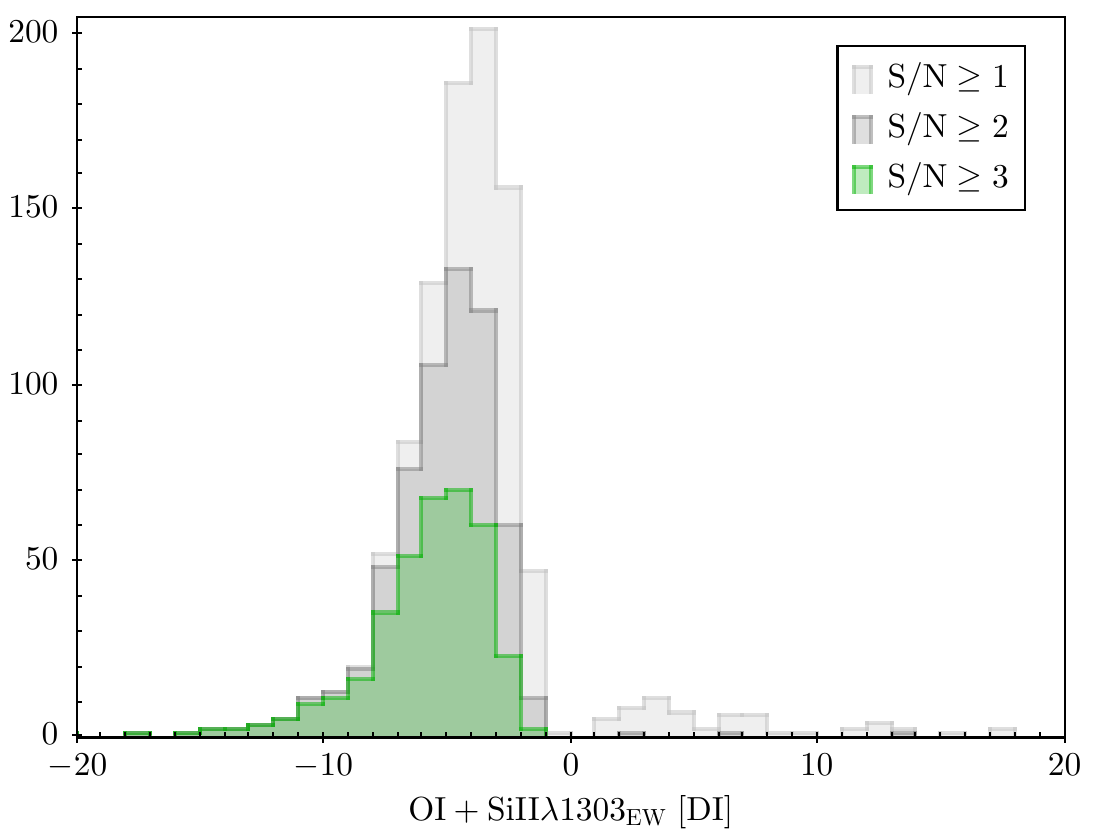}
        \includegraphics[scale=0.47]{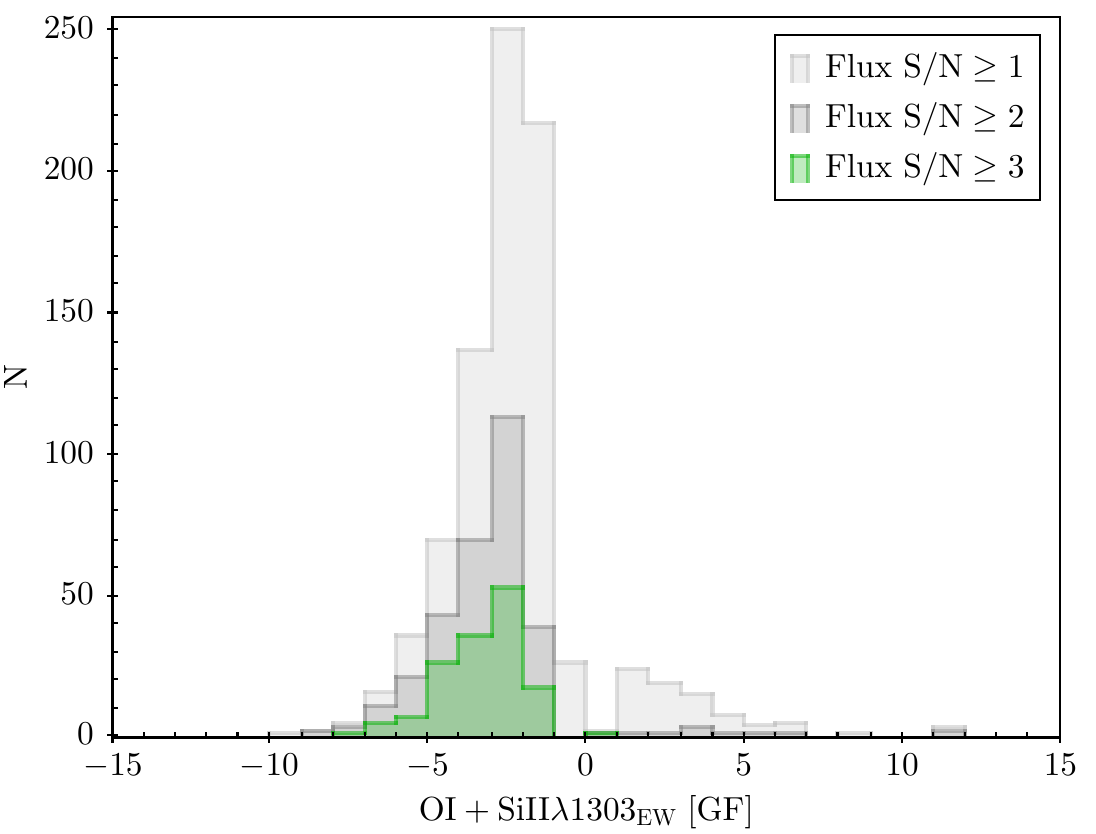}
        \caption{Distributions of the D4000 break and the EW of some notable lines. In each panel we show the distributions at S/N$\geq$1 (light grey) and S/N$\geq$3 (green). For the lines from the 'Gaussian fit' catalogues, the cut is in S/N flux.
        \emph{Top left}: Ly$\alpha$ EW (direct integration).    
        \emph{Top right}: D4000 (direct integration), 
        \emph{Middle left}: CIII]$\lambda$1909\AA~ EW (Gaussian fit). 
        \emph{Middle right}: [OII]$\lambda$3727\AA~ EW (Gaussian fit). 
        \emph{Bottom left}: OI+SiII$\lambda$1303\AA~ EW (direct integration). 
        \emph{Bottom right}: OI+SiII$\lambda$1303\AA~ EW (Gaussian fit).
        In the last two panels, we also show  the distribution at S/N$\geq$2 (dark grey). 
        } 
        \label{hist}
        \end{figure*}

We have produced a total of four catalogues: two (one for each field) for the Gaussian fit measurements performed with \texttt{slinefit} and two (again, one for each field) for the direct integration measurements performed with \texttt{pylick} plus Ly$\alpha$ following the \citet{kornei2010} method.
The contents of the catalogues are summarised in Table \ref{col}, while in Fig. \ref{hist} we show the distributions of the EW of some notable lines and, for the passive galaxies' sample at $z<2$, the D4000 break.
As already mentioned, we computed spectral properties only for galaxies with a reliable redshift, namely those with a quality flag = 2,3,4,9.
In the \texttt{slinefit} catalogue we have not included  the measurements for {the three} BLAGN whose emission line fits require two components\footnote{There are entries in the catalogue for these objects, but all cells in the table were set to -99.0.}. 
Dedicated spectral measurements for these objects will be presented in Bongiorno et al. (in prep.).

\begin{figure}
        \centering
        \includegraphics[scale=0.47]{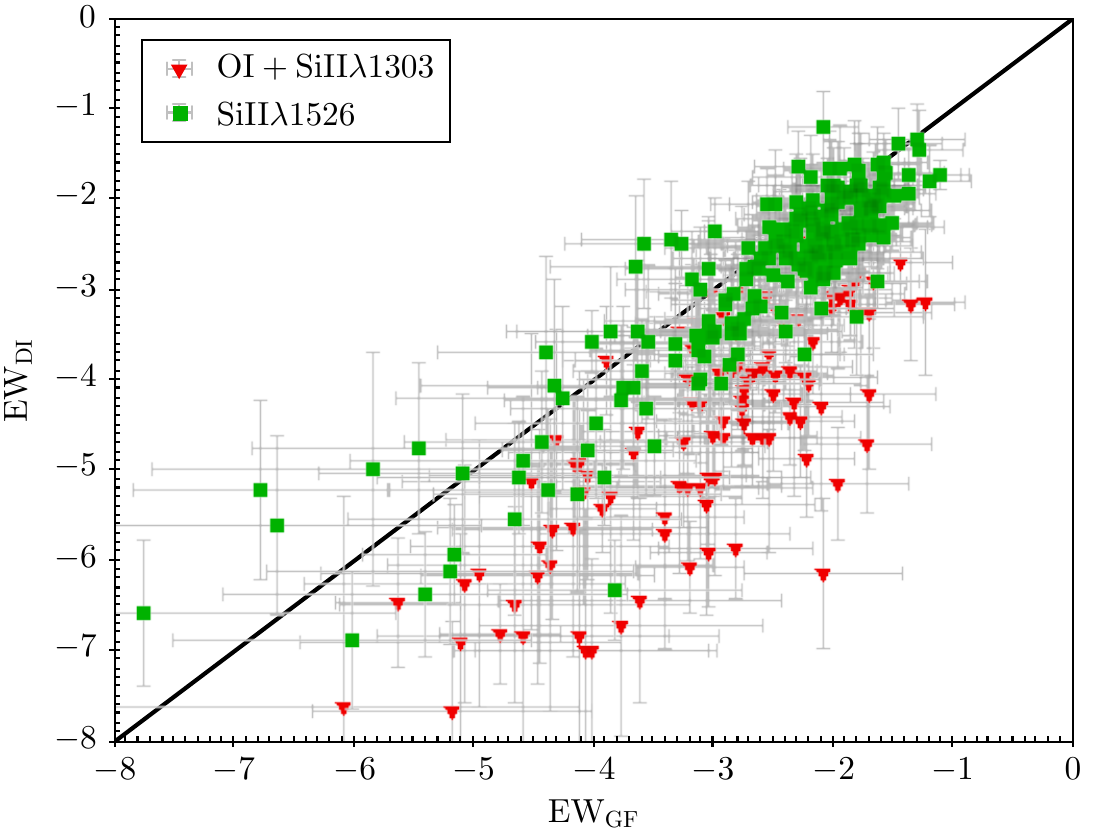}
        \caption{Examples of EW comparison between Gaussian fit and direct integration methods for a single line (SiII$\lambda$1526\AA; red triangles) and an unresolved group of lines (OI+SiII$\lambda$1303\AA; green squares). Only measurements at S/N$\geq$3 are shown. The 1-to-1 relation is also indicated in black.
        } 
        \label{gf_vs_di}
        \end{figure}

The EW of some lines was measured using both the Gaussian fit and direct integration methods. 
The agreement between the two measurements is very good in the case of single lines, as shown in Fig. \ref{gf_vs_di} for the SiII$\lambda$1526\AA, as an example: on average, the linear correlation coefficient is $r_{xy}\gtrsim{0.9}$ and the root-mean-square error (RMSE) is $\sim$0.5-0.6\AA. 
A systematic small offset of $\lesssim$0.5\AA~ is attributable to the different ways of determining the continuum level in the two methods.
In the case of unresolved groups of lines, where a single-Gaussian model was assumed (e.g. OI+SiII$\lambda$1303\AA), the correlation coefficient between the two methods is still high ($r_{xy}\gtrsim{0.8}$, on average, with an RMSE of $\sim$0.7-0.8\AA), but the Gaussian fit tends to systematically underestimate the flux, more than what would be expected by accounting only for the differences in the continuum.
The EW ratio between the two methods is on average between 0.6 and 0.8, depending on the group of lines.

\begin{figure*}
        \centering
        \includegraphics[scale=0.47]{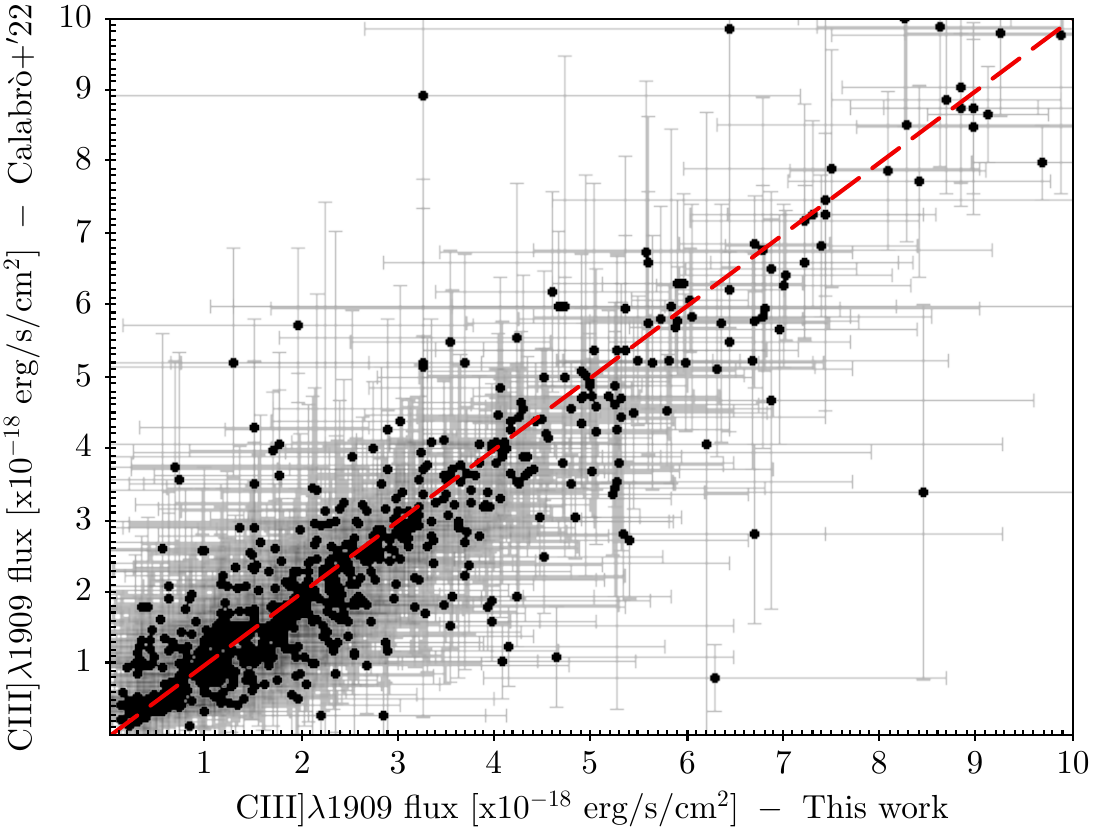}
        \includegraphics[scale=0.47]{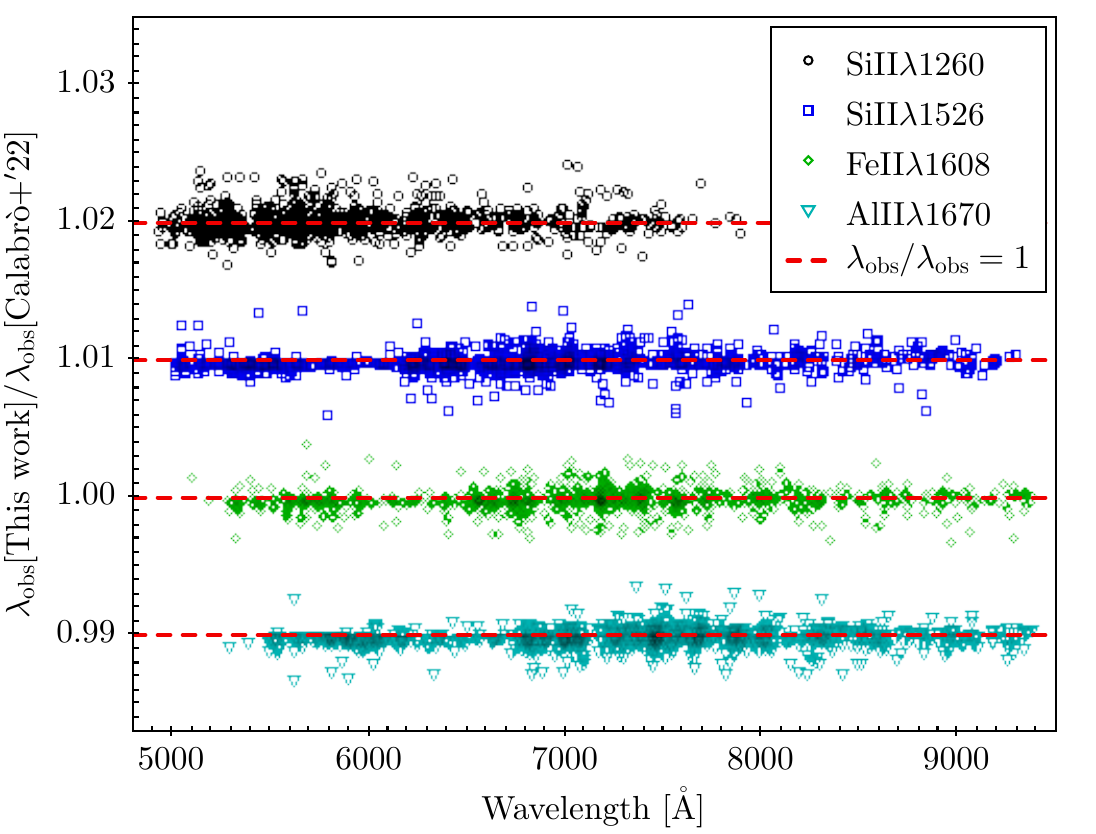}
        \includegraphics[scale=0.47]{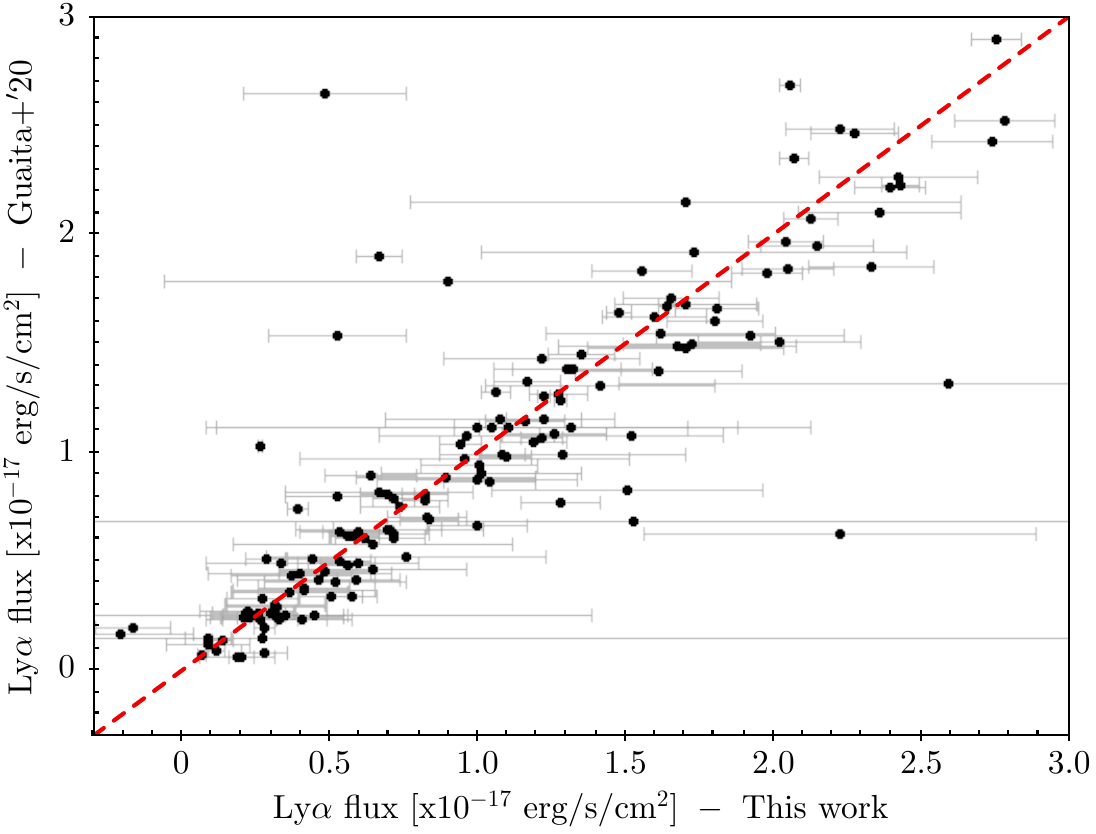}
        \includegraphics[scale=0.47]{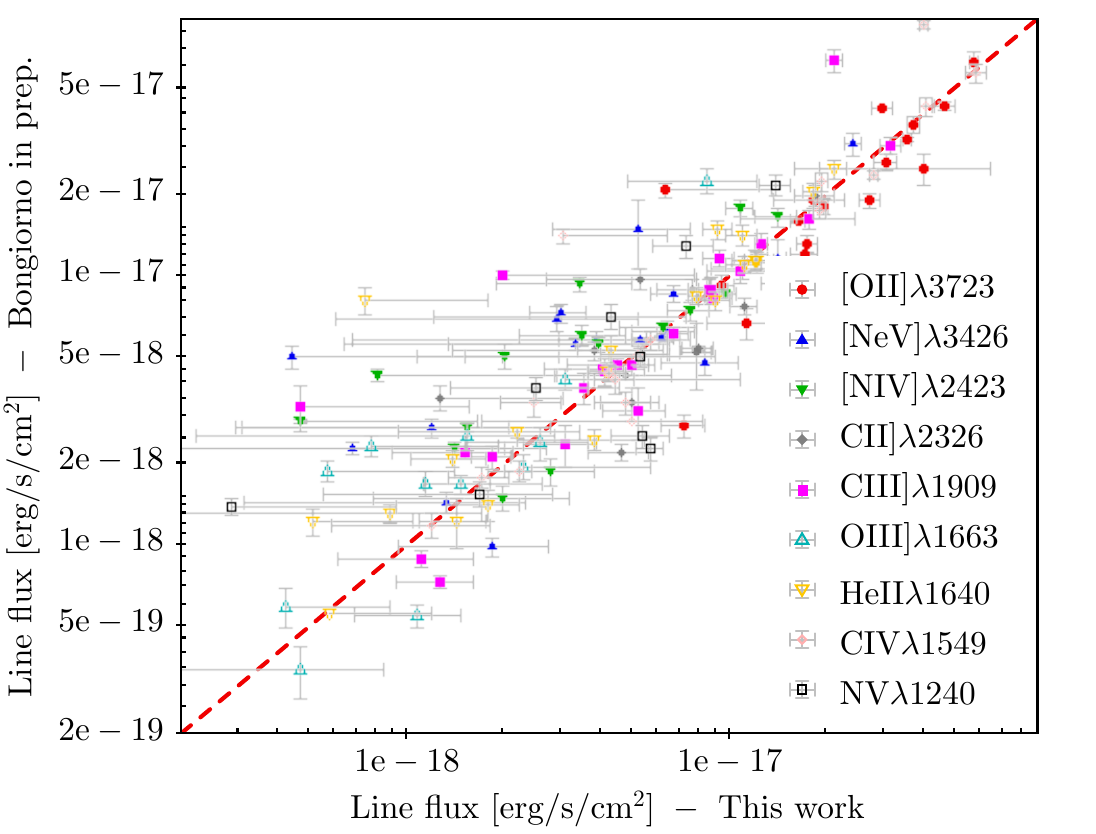}
        \includegraphics[scale=0.47]{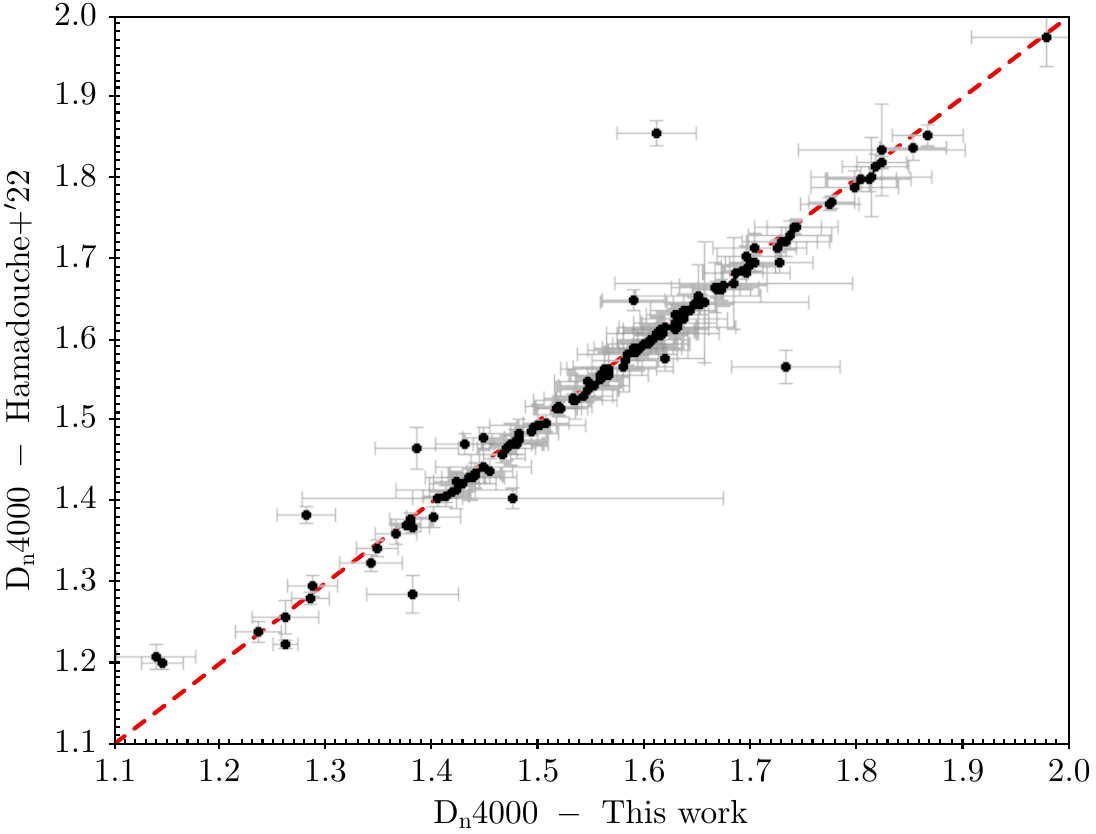}
        \caption{Comparison between the measurements presented in this work and previously published VANDELS results.
        In all plots we also show the one-to-one relation (dashed red line).
        \emph{Top left}: CIII]$\lambda$1909\AA~ flux from \citet{calabro2022b} (Gaussian fit). 
        \emph{Top right}: interstellar medium absorption line centroids from \citet{calabro2022b} (Gaussian fit; the points for the different ions have been shifted by 0.01 for visualisation purposes).
        \emph{Middle left}: Ly$\alpha$ flux from \citet{guaita2020} (Gaussian fit; no error was available for these measurements).
        \emph{Middle right}: AGN emission line flux from Bongiorno et al. (in prep.) (Gaussian fit).
        \emph{Bottom}: D$_{n}$4000 from \citet{hamadouche2022} (direct integration).
        } 
        \label{comp}
        \end{figure*}

Finally, as an additional validation, in Fig. \ref{comp} we compare subsets of measurements from our catalogues to independent and previously published measurements performed with different codes and methods by VANDELS team members.
For the Gaussian fit catalogue, we compared our measurements of the CIII]$\lambda$1909\AA~ flux and of the centroids of four absorption lines to those by \citet{calabro2022b}, which were obtained by fitting each line profile with a Gaussian function using the Python version of the \texttt{MPFIT} routine \citep{markwardt2009}. 
The continuum was parameterised as a straight line and fitted simultaneously with the lines.
We find a good agreement between the two sets of measurements (no S/N cut applied): for the CIII]$\lambda$1909\AA~ fluxes, the linear correlation coefficient is $r_{xy}\sim{0.9}$ and the RMSE$\sim$0.8$\times$10$^{-18}$erg s$^{-1}$ cm$^{2}$, with no significant offset with respect to the 1-to-1 relation. On the other hand, for the absorption lines' centroids $r_{xy}\sim{1.0}$ and RMSE$\sim$5.0$\times$10$^{-4}$\AA.

We checked the Ly$\alpha$ flux measurements from \citet{guaita2020}, which were obtained with a custom code based on the \emph{optimize.leastsq} Python function (see also \citet{guaita2017}), by fitting the lines with a Gaussian profile and assuming a linear continuum. 
\citet{guaita2020} provide two sets of measurements: one assuming a symmetric Gaussian profile, the other using a skewed Gaussian function. 
For our exercise, we took the former set (i.e. symmetric Gaussian) and limited the comparison to the galaxies with a goodness-of-fit flag for Ly$\alpha$ equal to 1 (see Sec. \ref{sec:gauss}).
 We also find in this case a good agreement between the two measurements, with an $r_{xy}\sim{0.9}$ and the RMSE$\sim$0.3$\times$10$^{-17}$erg s$^{-1}$ cm$^{2}$.

We compared our flux measurements for different emission lines in the VANDELS AGN sample (excluding BLAGN) to the ones obtained with a custom Python code from Bongiorno et al. (in prep.): the line fluxes are the mean of a Gaussian and a Lorentzian profile fit, plus a polynomial continuum.
The $r_{xy}$ and RMSE range from 0.7 to 0.9 and from 2.0$\times$10$^{-17}$ to 4.0$\times$10$^{-17}$erg s$^{-1}$ cm$^{2}$, respectively, depending on the line.
Finally, we checked the D$_{n}$4000 break in the VANDELS subsample of quiescent galaxies from our direct integration catalogue against the independent measurements by \citet{hamadouche2022} and we found an excellent agreement ($r_{xy}\sim{1.0}$ and RMSE$\sim$3.0$\times$10$^{-2}$).

\section{Summary}\label{sec:summary}
In this paper, we present the public release of the spectroscopic measurements of the VANDELS survey \citep{pentericci2018,mclure2018,garilli2021}.
We built two catalogues: one containing line properties from Gaussian fit measurements performed with the \texttt{slinefit} code, the other including line indices and continuum breaks measured with the \texttt{pylick} code, plus Ly$\alpha$ EWs following the \citet{kornei2010} method.
We created a set of mock spectra to mimic observed VANDELS sources in order to validate the \texttt{slinefit} code, while the \texttt{pylick} code was already tested in a previous work \citep{borghi2021}. 
As a further check of the accuracy of our catalogues, we compared subsets of measurements to previous results obtained with different codes and methods.
We have also found that the error spectra included in the VANDELS data release underestimate the noise level when compared to the r.m.s. of the object spectra and computed a correction that we provide in the catalogues. 
The full spectroscopic catalogues, together with the spectra, redshift catalogues, complementary photometric information, and SED fitting derived quantities, are publicly available from the VANDELS survey database\footnote{\url{http://vandels.inaf.it}} and at the CDS\footnote{via anonymous ftp to \url{cdsarc.cds.unistra.fr} (130.79.128.5) or via \url{https://cdsarc.cds.unistra.fr/cgi-bin/qcat?J/A+A/}}.

\newpage
\begin{acknowledgements}
        This paper is dedicated to the memory of Olivier Le F{\`e}vre.
        We would like to thank the anonymous referee for their constructive comments.
        The VANDELS Data Release 4 (DR4), including the catalogues presented in this papers, is publicly available and can be accessed using the VANDELS database at \url{http://vandels.inaf.it/dr4.html}, or through the ESO archives. 
        The data published in this paper have been obtained using the pandora.ez software developed by INAF IASF-Milano. 
        MT, LPoz and ACim acknowledge the support from grant PRIN MIUR 2017 20173ML3WW$\_$001.
        MT acknowledges the use of computational resources from the parallel computing cluster of the Open Physics Hub (\url{https://site.unibo.it/openphysicshub/en}) at the Physics and Astronomy Department in Bologna.
        ACim and MMor acknowledge the grants ASI n.I/023/12/0 and ASI n.2018-23-HH.0.
        Mmor acknowledges support from MIUR, PRIN 2017 (grant 20179ZF5KS)
        LPoz acknowledges the support from “fondi premiali” MITiC (MIning The Cosmos Big Data and Innovative Italian Technology for Frontier Astrophysics and Cosmology).
        LPen, ACal and MC acknowledge support from the Mainstream Grant VANDELS.
        The Cosmic Dawn Center (DAWN) is funded by the Danish National Research Foundation under grant No.140.
        JPUF acknowledges support from the Carlsberg Foundation.
        RA acknowledges support from ANID Fondecyt Regular 1202007.
        ACC thanks the Leverhulme Trust for their support via a Leverhulme Early Career Fellowship.
        MLH acknowledges the support of the UK Science and Technology Facilities Council.
        ASL acknowledges support from Swiss National Science Foundation. 

\end{acknowledgements}

\bibliographystyle{aa}
\bibliography{references} 

\end{document}